%% file: AABuGuS.tex
\documentclass[a4paper]{JHEP3}

\usepackage{
graphicx
,epsfig
,ifthen
,amsmath
}

\newcommand{\dmsk}{Der\-m\'{i}\-\v{s}ek}

\newcommand{\beqn}{\begin{eqnarray}}
\newcommand{\eeqn}{\end{eqnarray}}
\def\sla#1{\setbox0=\hbox{$#1$}\dimen0=\wd0
      \setbox1=\hbox{/} \dimen1=\wd1 \ifdim\dimen0>\dimen1
      \rlap{\hbox to \dimen0{\hfil/\hfil}} #1                        \else
      \rlap{\hbox to \dimen1{\hfil$#1$\hfil}}
      /   \fi}

\newcommand{\nn}{\nonumber}
\newcommand{\ov}{\overline}
\newcommand{\al}{\alpha}
\newcommand{\be}{\beta}

\newcommand{\eps}{\epsilon}

\newcommand{\la}{\lambda}

\newcommand{\mc}{\mathcal}

\newcommand{\ie}{{\em i.e.} }
\newcommand{\egv}{{\em e.g.}, }
\newcommand{\iev}{{\em i.e.}, }

\newcommand{\D}{\Delta}

\title{Challenging SO(10) SUSY GUTs 
with family symmetries through FCNC processes}

\author{Michaela~E.~Albrecht, Wolfgang~Altmannshofer,
Andrzej~J.~Buras, Diego~Guadagnoli, and David~M.~Straub\\ 
Physik-Department, Technische Universit\"at M\"unchen,\\
D-85748 Garching, Germany}

\abstract{We perform a detailed analysis of the SO(10) SUSY GUT model 
with $D_3$ family symmetry of \dmsk\ and Raby (DR). The model is specified in terms of
24 parameters and predicts, as a function of them, the whole MSSM set of parameters at low energy scales. 
Concerning the SM subset of such parameters, the model is able to give a satisfactory description of the quark and 
lepton masses, of the PMNS matrix and of the CKM matrix. We perform a global fit to the
model, including flavour changing neutral current (FCNC) processes $B_{s} \to \mu^+ \mu^-$, $B \to X_s \gamma$, 
$B \to X_s \ell^+ \ell^-$ and the $B_{d,s} - \ov B_{d,s}$ mass differences 
$\D M_{d,s}$ as well as the flavour changing (FC) process $B^+ \to \tau^+ \nu$. These observables provide at present the most sensitive probe of the SUSY mass spectrum 
and couplings predicted by the model. Our analysis demonstrates that the simultaneous
description of the FC observables in question represents a serious challenge for the DR model,
unless the masses of the scalars are moved to regions which are problematic from the point of view of naturalness and 
probably beyond the reach of the LHC. We emphasize that this problem could be a general feature of SUSY GUT models with 
third generation Yukawa unification and weak-scale minimal flavour violation.}
\preprint{TUM-HEP-671/07}

\begin{document}

\section{Introduction} \label{sec:intro}

A well-known problem of supersymmetry (SUSY) at the electroweak (EW) scale is 
its proliferation of parameters, arising if one keeps the most general allowed terms 
in the soft sector. In absence of the latter, SUSY is not phenomenologically viable, while
in presence of a most general soft sector, SUSY largely loses its predictivity, due 
to the bulkiness of the parameter space.

Since SUSY goes often together with the idea of Grand Unification, due to 
the tantalizing observation of gauge coupling unification after MSSM running, 
a sensible way-out to the above problem is to take a `top-down' approach.
In this case one starts with the theory at the GUT scale -- with a far simpler 
parameter space than the MSSM one -- and runs all the parameters downwards 
through Renormalization Group Equations (RGEs). The latter then dynamically 
generate all the mass splittings for the soft terms at the EW scale, 
which, in the low-energy MSSM without GUT, are treated instead as free parameters.

As an argument in favour of a top-down approach to the MSSM, it can be noted that predictive 
SUSY GUT models need typically two ingredients at the GUT scale. The first one is the choice of 
a specific GUT gauge symmetry. Second, one has to fix the high-scale soft sector with minimal 
assumptions on the parametric dependence, which seems justified due to a higher amount of symmetry 
of the theory at this high scale. A further restriction on the number of parameters and hence more 
predictability of the model can be obtained by introducing additional family symmetries.
Once these different parts of the model are specified, the theoretical and computational tools 
available today allow for a very controlled theoretical error on the model parameters at low-energy, 
in spite of the `long running' from the GUT scale and the presence of mass thresholds.

Concerning SUSY GUT models present on the market, while it is easy to construct models reproducing 
the gauge sector of the MSSM at low-energy, it is far more challenging to find models also correctly 
describing `flavour patterns' such as quark and lepton masses and the CKM and PMNS matrices.

One such notable model has been proposed by \dmsk\ and Raby in \cite{DR05}. It is an
SO(10) SUSY GUT, augmented with a simple family symmetry, conserving R-parity at low-energy. 
As shown in \cite{DR05}, this model is able to successfully fit
all the parameters of the SM. In particular, by using some observables basically
unaffected by SUSY contributions, it can reproduce the CKM matrix entries. Finally, it
also describes the known parameters in the neutrino sector. 
In Ref. \cite{DR06}, the same model was also extensively studied in the {\em lepton} flavour 
sector, thereby providing a number of signatures of the model, which should at least in part 
be tested by forthcoming experiments.

In view of the above mentioned remarkable performance of the DR model in describing low-energy
observables, it is interesting to have a closer look at its SUSY spectrum, with the aim of
testing, \egv the predicted mass hierarchies. Since SUSY particles have not yet been observed, 
such task can only be accomplished by analyzing implied loop effects in flavour changing 
neutral current (FCNC) processes. 
In fact, the specific mass patterns of gauginos, up- and down-squarks and Higgs multiplets
predicted by the model, do affect measured quark FCNC observables in a peculiar way,
and if such observables are well controlled, the pattern of implied corrections is
testable. Fortunately, we have today a whole host of such observables, which have the virtue 
of being at the same time precisely measured and accurately calculated within the MSSM.

The aim of the present paper is then an in-depth test of the model in the light of all the
most accurate information presently available on quark FCNCs. The flavour sector is often
overlooked in first analyses of new physics (NP) models, due to its vastness and the necessity of sometimes
involved calculations. In the context of the present model, we show that the flavour
sector has nonetheless enough sensitivity to the details of the SUSY spectrum, to represent 
a discriminating test for the model itself.

\medskip

Our paper is organized as follows. In Section \ref{sec:model} we give a brief overview of
the DR model, focusing on those ingredients that are most relevant for our purposes. In
Section \ref{sec:procedure} we then outline the procedure to connect the GUT scale model
with low-energy observables. Such procedure is well-known, but often obscured by the
assumptions made on the running and the mass thresholds. 
Section \ref{sec:obs} goes then in more detail on the determination of
masses and couplings, through consideration of low-energy threshold effects. In Section
\ref{sec:FCNC} we then present the collection of FCNC observables we use for the
purpose of our paper. The emphasis here is on presenting simplified expressions, in order to
provide an intuitive picture of the main effects, with refined formulae only used in the
actual numerical analysis. In the light of such intuitive expressions, Section
\ref{sec:picture} then discusses the general pattern featured by the corresponding FCNC
observables within the DR model. An extensive analysis of all such features upon variation
of the DR model parameters is then presented in Section \ref{sec:fit}. Finally, Section
\ref{sec:conclusions} is devoted to our conclusions.

\section{The model} \label{sec:model}

The DR model \cite{DR05,DR-D3symmetry} is a supersymmetric SO(10) Grand Unified Theory
with an additional $D_3 \times [U(1) \times Z_2 \times Z_3]$ family symmetry.  

The above symmetry group fixes the following structure for the superpotential
\beqn
W = W_f + W_\nu ~, \nonumber
\eeqn
with
\beqn
\label{Wcf}
W_f &=& \textbf{16}_3 \, \textbf{10}\, \textbf{16}_3+\textbf{16}_a \, \textbf{10}\, \chi_a \nn \\
&+&\bar{\chi}_a (M_{\chi}\, \chi_a+\textbf{45} \,\frac{\phi_a}{\hat{M}}\, \textbf{16}_3
+\textbf{45} \,\frac{\tilde{\phi}_a}{\hat{M}} \, \textbf{16}_a+A\, \textbf{16}_a)~,\\
W_\nu &=&\ov{\textbf{16}} (\lambda_2\, N_a\,\textbf{16}_a+\lambda_3\,N_3\,\textbf{16}_3) 
+\frac{1}{2} (S_a\,N_a\,N_a+S_3\,N_3\, N_3)~.
\eeqn

The first two families of quarks and leptons are contained in the superfield
$\textbf{16}_a,\:a=1,2$, which transforms under SO(10)$\times D_3$ as $(\textbf{16}, \textbf{$2_A$})$, 
whereas the third family in $\textbf{16}_3$ transforms as $ (\textbf{16},\textbf{$1_B$})$. 
The two MSSM Higgs doublets $H_{u}$ and $H_d$ are contained in a $\textbf{10}$. As can be seen from the 
first term on the right-hand side 
of (\ref{Wcf}), Yukawa unification $\lambda_t=\lambda_b=\lambda_\tau=\lambda_{\nu_\tau}$ at $M_G$ is 
obtained {\em only} for the third generation, which is directly coupled to the Higgs $\textbf{10}$ 
representation. This immediately implies large $\tan\beta \approx 50$ at low energies.

The effective Yukawa couplings of the first and second generation fermions are instead generated via the 
Froggatt-Nielsen mechanism \cite{FroggattNielsen} as follows. Additional fields are introduced, \ie the 
$\textbf{45}$ which is an adjoint of SO(10), the SO(10) singlet flavon fields $\phi^a, \tilde {\phi^a}, A$ 
and the Froggatt-Nielsen states $\chi_a, \bar{\chi}_a$. 
The latter transform as a $(\textbf{16},\textbf{$2_A$})$ and a $(\ov{\textbf{16}},\textbf{$2_A$})$,
respectively, and receive masses of O$(M_{G})$ as $M_\chi$ acquires an SO(10) breaking VEV. 
Once they are integrated out, they give rise to effective mass operators which, together with the 
VEVs of the flavon fields, create the Yukawa couplings for the first two generations.
This mechanism breaks systematically the full flavour symmetry and produces the right 
mass hierarchies among the fermions.

The obtained Yukawa matrices for up-quarks, down-quarks, charged leptons and neutrinos are
\beqn
&Y_u=\left(
\begin{array}{ccc}
0 & \varepsilon' \,\rho & -\varepsilon\,\xi \\
-\varepsilon' \, \rho & \tilde{\varepsilon} \,\rho & -\varepsilon \\ 
\varepsilon\,\xi & \varepsilon & 1
\end{array}
\right)\,\lambda~,&~~
Y_d=\left(
\begin{array}{ccc}
0 & \varepsilon'  & -\varepsilon\,\xi\,\sigma \\
-\varepsilon'  & \tilde{\varepsilon}  & -\varepsilon\,\sigma \\ 
\varepsilon\,\xi & \varepsilon & 1
\end{array}
\right)\,\lambda~,\nn \\
&Y_e=\left(
\begin{array}{ccc}
0 & -\varepsilon'  & 3\,\varepsilon\,\xi \\
\varepsilon'  & 3\,\tilde{\varepsilon}  & 3\,\varepsilon \\ 
-3\,\varepsilon\,\xi\,\sigma & -3\,\varepsilon\,\sigma & 1
\end{array}
\right)\,\lambda~,&~~
Y_\nu=\left(
\begin{array}{ccc}
0 & -\varepsilon' \,\omega & \frac{3}{2}\,\varepsilon\,\xi \,\omega\\
\varepsilon'  \,\omega& 3\,\tilde{\varepsilon}\,\omega  
& \frac{3}{2}\,\varepsilon\,\omega \\ -3\,\varepsilon\,\xi\,\sigma 
& -3\,\varepsilon\,\sigma & 1
\end{array}
\right)\,\lambda~.
\label{Y-textures}
\eeqn
From eqs. (\ref{Y-textures}) one can see that the flavour hierarchies in the Yukawa
couplings are encoded in terms of the four complex parameters $\rho, \sigma, \tilde \varepsilon, \xi$ 
and the additional real ones $\varepsilon, \varepsilon', \lambda$.

In order to avoid neutrino masses of the order of the other fermion masses,
one invokes the type-I see-saw mechanism \cite{Minkowski,see-saw1,see-saw2,see-saw3}. In particular, three SO(10) singlet Majorana fermion fields 
$N_a, N_3$ $(a=1,2)$ are introduced via the contribution of $\frac{1}{2}\, (S_a\,N_a \, N_a+S_3 \,N_3\,N_3)$ to the superpotential.

The mass term $\frac{1}{2}\,N\,M_N\,N$ is produced when the flavon fields acquire VEVs 
$\langle S_a\rangle=M_{N_a}$ and $\langle S_3\rangle=M_{N_3}$. Together with a $\ov{\textbf{16}}$ Higgs one is allowed to 
introduce the interaction terms $\ov{\textbf{16}} \,(\lambda_2 \, N_a\, \textbf{16}_a+\lambda_3 \, N_3\, \textbf{16}_3)$, 
which in turn generate a mixing matrix $V$ between the right-handed neutrinos and the additional singlets ($\ov{\nu}\, V \, N$), 
when the $\ov{\textbf{16}}$ acquires an SO(10) breaking VEV $\langle \ov{\textbf{16}} \rangle_{\ov{\nu}} =v_{16}$. The resulting 
effective right-handed neutrino mass terms read 
\beqn
 W_N= \ov {\nu}\, V\,N+\frac{1}{2}\,N\,M_N\,N~,
\eeqn
\beqn
V=v_{16}\left(\begin{array}{ccc}
0 & \lambda_2 & 0 \\
\lambda_2 & 0 & 0 \\ 
0 & 0 & \lambda_3 
\end{array}\right)~,~~~~
M_N={\rm diag}(M_{N_1},M_{N_2},M_{N_3})~.
\eeqn
Diagonalization leads to the effective right-handed neutrino Majorana mass
\beqn
M_R = - V \, M_N^{-1}\, V^T \equiv - {\rm diag}(M_{R_1},M_{R_2},M_{R_3}) ~.
\eeqn
By integrating out the EW singlets $\overline{\nu}$ and $N$, which both receive
GUT scale masses, one ends up with the light neutrino mass matrix at the EW scale given by the
usual see-saw formula
\beqn
\mc M = m_\nu\, M_R^{-1}\,m_\nu^T~.
\eeqn

\section{Basic procedure} \label{sec:procedure}

In this section, we describe the procedure which, from the specification of the model at
the GUT scale, leads to the MSSM mass spectrum and observables at the EW scale.

\subsection{Step 1: Parameters of the model} \label{sec:pars}

\subsubsection*{Gauge coupling sector}

We choose as three parameters the unification scale $M_G$, the gauge coupling $\al_G$ defined through
\beqn
\alpha_G \equiv \alpha_1(M_G) = \alpha_2(M_G)~,
\eeqn
and the threshold correction $\epsilon_3$ defined through
\beqn
\al_3(M_G) = \al_G (1+\epsilon_3)~.
\eeqn
The threshold correction $\epsilon_3$ helps to obtain the right value of $\al_3(M_Z)$.

\subsubsection*{SUSY sector}
We have the following set of soft SUSY breaking parameters: 
a universal sfermion mass $m_{16}$, a universal gaugino mass $M_{1/2}$, a universal 
trilinear coupling parameter $A_0$, and the Higgs mass parameters $m_{H_u}$ and 
$m_{H_d}$. It is well-known that at large $\tan\beta$, EWSB is easier to achieve by 
allowing the soft SUSY breaking Higgs mass parameters to be split. This also has the 
consequence that the absolute value of $\mu$ is not fixed by EWSB, as is the case in the 
CMSSM\footnote{
Very interesting combined analyses in the framework of the CMSSM have been 
performed in \cite{ATR,RAT,RAT2}. In \cite{Ellis2007}, also non-universal Higgs masses were considered, but only with vanishing $A_0$.
}, but is instead a free parameter.

\subsubsection*{Yukawa matrices}
The Yukawa matrices for up- and down-type quarks, charged leptons and neutrinos are
parameterized as given in (\ref{Y-textures}), with $\varepsilon, \varepsilon', \lambda$
being real and $\rho, \sigma, \tilde \varepsilon, \xi$ complex parameters.

\subsubsection*{Right-handed neutrinos}
The diagonal right-handed neutrino mass matrix is chosen to be real, 
which amounts to three parameters $M_{R_i}$, with $i=1,2,3$.

\subsubsection*{Weak-scale parameters}
In addition to the above GUT-scale parameters, the SUSY parameters $\mu$ and $\tan\beta$ 
have also to be specified as an input at the weak scale.

\bigskip
\TABLE{
\begin{tabular}{|lcc|}
\hline
Sector & \# & Parameters \\
\hline \hline
gauge & 3 & $\alpha_G$, $M_G$, $\eps_3$, \\
SUSY (GUT scale) & 5 & $m_{16}$, $M_{1/2}$, $A_0$, $m_{H_u}$, $m_{H_d}$, \\
textures & 11 & $\eps$, $\eps'$, $\la$, $\rho$, $\sigma$, $\tilde \eps$, $\xi$, \\
neutrino & 3 & $M_{R_1}$, $M_{R_2}$, $M_{R_3}$, \\
SUSY (EW scale) & 2 & $\tan \beta$, $\mu$ \\
\hline \hline
\end{tabular}
\caption{Parameters in the DR model.}
\label{tab:parameters}
}

The {\em total} number of input parameters, listed in this step and summarized in Table
\ref{tab:parameters}, is therefore 24. Once these parameters are fixed, the MSSM couplings 
as well as its {\em whole} mass spectrum (which includes the SM part) can be predicted at energies 
below $M_G$, in particular at $M_Z$ or any lower scale relevant for FCNC processes. Of course 
a subset of the observables amenable to prediction is {\em used} to fix the model parameters.

The procedure to fix the above parameters will be addressed in detail in Section \ref{sec:fit}.
Such procedure uses low-energy observables, and for this reason we need to evolve the
fundamental parameters of the theory to $M_Z$ or below, using their RGEs. The evolution to low 
energy is performed according to a procedure we will now describe.

\subsection{Step 2: RG evolution} \label{sec:RGE}

\subsubsection*{\boldmath $M_{\rm GUT} >$ scale $> M_{R_i}$}
The RGEs of the MSSM with right-handed neutrinos \cite{Hisano95, AKLR05, Petcov03} 
are used to run the gauge couplings, Yukawa couplings, all soft SUSY breaking 
parameters and the right-handed neutrino mass matrix $M_R$ down to the mass of the 
heaviest right-handed neutrino. At this threshold, one must rediagonalize $M_R$ and  
integrate out one neutrino following \cite{AKLR02}. This generates the effective 
dimension-five neutrino mass operator. The two remaining thresholds are treated 
accordingly, with the number of neutrinos reduced by one each time.
In practical calculations, this procedure turns out to be computationally demanding.
We have thus followed the approach of integrating out all the right-handed neutrinos at a single intermediate
threshold, corresponding to the mass of the lightest of them. We have then checked that 
this approximate treatment does not have any relevant impact on either the determination 
of the GUT-scale parameters or the low-energy predictions of the model.

\subsubsection*{\boldmath $M_{R_i} >$ scale $> M_Z$}
Having constructed the effective theory without right-handed neutrinos at the scale where
they are integrated out, we use MSSM RGEs \cite{MartinVaughn} to run the gauge couplings, 
Yukawa matrices, soft SUSY breaking parameters and the Wilson coefficient of the neutrino mass 
operator down to the scale $M_Z$. We use two-loop RGEs for dimensionless and one-loop 
for dimensionful parameters.

\bigskip

So, starting from the fundamental parameters of Step 1, and performing RGE evolution
through Step 2, one has now the whole set of parameters of the MSSM fixed at $M_Z$ or 
below, to tree-level. However, for many quantities, such as gauge couplings, some SUSY masses, quark 
masses, CKM matrix and the Higgs sector, a one-loop determination turns out to be
mandatory, for different reasons. These issues will be discussed in the next section.

Here we would like to stress a further property of the low-scale MSSM parameters implied by the DR model. 
The off-diagonal entries in the squark mass matrices are generated radiatively by the Yukawa couplings;
in addition the quark sector features only one CP phase\footnote{Putting additional phases
in the right-handed neutrino mass matrix would communicate corresponding phases to the low-energy neutrino 
sector, without affecting, to a very good approximation, the quark sector.}. These two facts allow to 
classify the low-energy MSSM resulting from the RG evolution of the DR model as belonging to the class 
of models with minimal flavour violation (MFV) \cite{MFV}.

With the above qualifications, using the numerically determined parameters of the theory,
we can then evaluate 
effective Hamiltonians for weak decays, in particular for FCNC and
CP-violating processes. Having these Hamiltonians at hand, it is straightforward to
evaluate the branching ratios for various low energy processes and to compare them with
experiments. The latter strategy, allowing to extensively test the model in the flavour
sector, will be dealt with in Sections \ref{sec:picture} and \ref{sec:fit}.

\section{One-loop improved determination of low-energy observables} \label{sec:obs}

With low-energy observables we denote all the physical quantities that can be directly
accessed experimentally and compared with predictions of the model. Among these
quantities, some will be used to fix the model parameters, through a fitting procedure, as
described in Section \ref{sec:fit}. The others will then be genuine predictions of the model. 

To obtain all physical observables beyond tree-level accuracy at the scale $M_Z$, 
one needs to include one-loop corrections in the relevant formulae defining the
observables themselves. 
The inclusion of such corrections is also advocated in the analyses \cite{DR05,DR06}. 
Below, we give some details of the procedure in the various cases.

\subsubsection*{SUSY masses}

The mass eigenstates for squarks, sleptons, charginos and neutralinos are
calculated at the weak scale from the tree-level mass matrices \cite{RosiekFR}. The gluino
pole mass is calculated at the one-loop level.

\subsubsection*{Gauge sector}
We calculate the gauge couplings $\alpha_s(M_Z)$ and $\alpha_{\rm em}$
by including the threshold corrections given in \cite{PBMZ}. 
We use tadpole corrections to the Higgs potential \cite{PBMZ,ChankowskiOSH} to obtain the 
one-loop VEV and include one-loop SUSY corrections to the Fermi constant measured 
in muon decay, $G_\mu$, as well as to the $W$ and $Z$ boson pole masses \cite{ChankowskiOSH}.

\subsubsection*{Higgs spectrum}
The Higgs spectrum is represented by the masses $M_{h_0}$, $M_{H_0}$, $M_{H^+}$, $M_A$ 
of the corresponding physical particles. In this model $H_0$, $H^+$ and $A$ are typically nearly
degenerate, with only the lightest Higgs mass $M_{h_0}$ lying at a substantially lower
value. The scale of the masses $M_{H_0}$, $M_{H^+}$, $M_A$ is set in a non-trivial way by the
interplay among $\mu, \tan \be, m_{H_{u,d}}$ and $m_{16}$ in the equations minimizing the
Higgs potential.

We calculate the pseudoscalar Higgs pole mass $M_A$ following \cite{PBMZ} and use it as
an input to {\tt FeynHiggs 2.5.1} \cite{FeynHiggs1,FeynHiggs2,FeynHiggs3,FeynHiggs4}, 
which accurately calculates the masses of the remaining Higgs mass eigenstates,
using Yukawa matrices and soft terms at the EW scale, which in turn are the result of 
the RG analysis. 

We explicitly note that $M_A$ (and with it the other heavy Higgs masses) is typically 
`pushed up' by the upper bound on the BR$(B_s \to \mu^+ \mu^-)$, so that the $M_A$ value
obtained with any given choice of the rest of the parameters can be considered as a lower
mass bound on the heavy Higgs spectrum.

\subsubsection*{Fermion masses and CKM matrix}

At large $\tan\beta$, quark masses undergo $\tan\beta$ enhanced corrections that have to be 
included to the tree-level determination, represented by the running Yukawa couplings at $M_Z$ \cite{CGNW}. 
These corrections also modify the relations between the original CKM matrix appearing in the MSSM 
Feynman rules and the effective CKM matrix measured in tree-level decays \cite{BRP,BCRSbig}.
We closely follow the line of argument of \cite{BCRSbig} and calculate the one-loop threshold 
corrections to quark and charged lepton mass matrices at $M_Z$, but take into account both 
SUSY and electroweak contributions.
After applying the threshold corrections, we use three-loop QCD and one-loop QED RGEs to run 
the five light quark and three charged lepton masses down to their respective scales.
 
The neutrino masses and the PMNS mixing matrix, on the other hand, are left at the tree-level.
In fact we find the threshold corrections to these quantities, discussed in \cite{ChankowskiWasowicz}, 
to be numerically negligible.

\section{Basic formulae for FC observables} \label{sec:FCNC}
In this section we collect formulae for various branching ratios that we will 
use in our numerical analysis. In certain cases we show only the leading contribution 
for large $\tan\beta$, in order to provide an intuitive picture of the behaviour. 
For example, in the case of $B_{s,d}\rightarrow \mu^{+} \mu^{-}$ decays, this corresponds 
to the contribution of Higgs penguins. In the actual numerical analysis we 
include all the relevant contributions, \iev besides the SM one, contributions from
charginos, charged  Higgses and gluinos. Neutralino contributions are generally 
negligible.

\subsection[$B_{s,d}\rightarrow \mu^{+} \mu^{-}$]
{\boldmath $B_{s,d}\rightarrow \mu^{+} \mu^{-}$} \label{sec:bsmumu}

In the SM, the usual $Z$-penguin and box diagrams result in strongly suppressed branching 
ratios that are sensitive functions of the weak decay constants $F_{B_{d}}$ and $F_{B_{s}}$. 
Eliminating this dependence with the help of the well-measured mass differences 
$\Delta M_{s,d}$ \cite{BurasPLB}, one finds \cite{BBGT}
\beqn
{\rm BR}(B_s\rightarrow \mu^{+} \mu^{-})_{{\rm SM}}&=&(3.37\pm 0.31)\times 10^{-9}~, \label{Bsmumu} \\
{\rm BR}(B_d\rightarrow \mu^{+} \mu^{-})_{{\rm SM}}&=&(1.02\pm 0.09)\times 10^{-10}~. \label{Bdmumu}
\eeqn
These values should be compared with the present 95\% C.L. upper bounds from CDF
\cite{CDF-bsmumu}
\beqn
{\rm BR}(B_s\rightarrow \mu^{+} \mu^{-})_{\rm exp} < 1.0 \times 10^{-7}~,~~~~
{\rm BR}(B_d\rightarrow \mu^{+} \mu^{-})_{\rm exp} < 3.0 \times 10^{-8}~, \label{bsmumu-exp}
\eeqn
that leave still a large room for NP contributions.

In the MSSM with large $\tan\beta$, the helicity suppression in 
(\ref{Bsmumu}) and (\ref{Bdmumu}) is lifted by Higgs-mediated neutral currents
\cite{gaur,BabuKolda}. Their contributions can be summarized by the approximate formula
\cite{BCRSbig,IsidoriRetico}
\beqn \label{bsmumuSUSY}
{\rm BR}(B_s\to\mu^+\mu^-)&\simeq&3.5\times 10^{-5}
\Bigg[\frac{\tan\beta}{50}\Bigg]^6
\Bigg[\frac{\tau_{B_s}}{1.5\ {\rm ps}}\Bigg]
  \Bigg[\frac{F_{B_s}}{230\ {\rm MeV}}\Bigg]^2 
  \Bigg[\frac{|V_{ts}|}{0.040}\Bigg]^2\nonumber\\
&\times&\frac{\ov m_t^4}{M^4_A} 
\frac{(16\pi^2\epsilon_Y)^2}
{(1+\tilde\epsilon_3\tan\beta)^2(1+\epsilon_0\tan\beta)^2}~,
\eeqn
where $\overline m_t\equiv m_t(\mu_t)$ and 
\beqn
\tilde\epsilon_3=\epsilon_0+y_{t}^{2} \epsilon_Y~,
\eeqn
with $\epsilon_0$ and $\epsilon_Y$ standing for gluino loop and chargino loop factors,
whose full expressions can be found in \cite{BCRSbig}.

With a similar expression for BR$(B_d\to\mu^+\mu^-)$, one also gets
\beqn \label{BR/BR}
\frac{{\rm BR}(B_d\to\mu^+\mu^-)}{{\rm BR}(B_s\to\mu^+\mu^-)} 
=\left[\frac{\tau_{B_d}}{\tau_{B_s}}\right]
 \left[\frac{F_{B_d}}{F_{B_s}}\right]^2 
 \left[\frac{|V_{td}|}{|V_{ts}|}\right]^2
\left[\frac{M_{B_d}}{M_{B_s}}\right]^5~,
\label{eq:bsmumu/bdmumu}
\eeqn
where non-leading contributions have been neglected.

Observing that the ratio (\ref{eq:bsmumu/bdmumu}) is roughly a factor of ten smaller than 
the corresponding ratio of experimental bounds (\ref{bsmumu-exp}),
it is clear that, in our framework, the current ${\rm BR}(B_d\to\mu^+\mu^-)$ constraint is completely 
marginal with respect to the $B_s$ counterpart, which is the only channel considered in the 
rest of the analysis.

We explicitly note that, in eq. (\ref{bsmumuSUSY}), as throughout the text, $V_{ti}$ denotes 
elements of the {\em physical} CKM matrix\footnote{In the notation of \cite{BCRSbig}, such 
matrix elements are written as $V_{ti}^{\rm eff}$.}, to be distinguished for large $\tan \beta$ from 
the corresponding matrix appearing at the Lagrangian level. The latter is denoted as `bare' 
since it does not yet include the large $\tan \be$-resummed effects. A similar comment applies 
to the quark masses. In actual numerical calculations, the differences between the physical and 
the `bare' parameters have been taken into proper account as discussed in section~\ref{sec:obs}.

At this stage, it suffices to state that $\epsilon_0$ and $\tilde\epsilon_3$ are at 
most O(10$^{-2}$). Still, with $\tan\beta \approx 50$, as characteristic for the DR model,
the $\tan\beta$-resummed corrections in the last factor in (\ref{bsmumuSUSY}) can be significant 
and, depending on the sign of $\mu$, can provide an additional 
enhancement of ${\rm BR}(B_s\to\mu^+\mu^-)$ or respectively some suppression relative to the 
leading behaviour $(\tan\beta)^{6}$. One finds in the full space of parameters considered
\beqn
{\rm sign}(\epsilon_0)={\rm sign}(\tilde\epsilon_3)={\rm sign}(\mu)
\eeqn
and more explicitly,
\beqn
\epsilon_0 &\approx& - \frac{2 \alpha_{s}}{3 \pi} \frac{\mu}{m_{\tilde g}} 
H_{2}(m^2_{\tilde b_1}/m_{\tilde g}^2, m^2_{\tilde b_2}/m_{\tilde g}^2)~, \\
\epsilon_Y &\approx& \frac{1}{16 \pi^{2}} \frac{A_{t}}{\mu} 
H_{2}(m^2_{\tilde t_1}/\mu^2,m^2_{\tilde t_2}/\mu^2)~,
\eeqn
where $m_{\tilde b_i}$ ($m_{\tilde t_i}$) are the masses of the $i^{\rm th}$ sbottom
(stop), $A_t$ is the soft SUSY breaking stop trilinear parameter\footnote{
Our sign convention for $A_t$ is such that the off-diagonal entry of the tree-level stop mass matrix 
reads $m_t(A_t-\mu\cot\beta)$. This agrees with the sign convention for $A_0$ in \cite{DR05,DR06} 
but disagrees with the convention used in \cite{RosiekFR,BCRSbig}.
\label{fn:A-sign}
}
and $m_{\tilde g}$ the gluino 
mass. The function $H_2$ is defined as \cite{BCRSbig} (see also \cite{CMW})
\beqn
H_2(x,y) = \frac{x \log x}{(1-x)(x-y)} + \frac{y \log y}{(1-y)(y-x)}~.
\eeqn

We emphasize that in the DR model the parameters entering $\epsilon_0$ and $\epsilon_Y$ 
are strongly correlated with each other and consequently the range of values which 
$\epsilon_0$ and $\epsilon_Y$ can take is significantly smaller than in the usual 
studies of the above formulae, that can be found in the literature.

We also emphasize that in the numerical analysis one can replace the above branching
ratio with the quantity \cite{BurasPLB}
\beqn \label{BR/DM}
\frac{{\rm BR}(B_s\to\mu^+\mu^-)}{\Delta M_s}~,
\eeqn
thereby eliminating $F_{B_{s}}$ that is still inaccurately known. The only hadronic 
uncertainties in the ratio in (\ref{BR/DM}) are present in the non-perturbative 
factors $B^i_s$, that enter $\Delta M_{s}$ only linearly and are better known from lattice 
calculations than the decay constant \cite{Bparams-lattice-DB}.

\subsection[$\Delta M_{s,d}$]{\boldmath $\Delta M_{s,d}$}

The mass differences in the $B_{s,d} - \ov B_{s,d}$ systems, $\Delta M_{s,d}$, consist, in
the MSSM at large $\tan \beta$, of the following contributions
\beqn \label{DMsformula}
\Delta M_{s,d} =
\Delta M_{s,d}^{\rm SM} + \Delta M_{s,d}^{H^{+}}
+ \Delta M_{s,d}^{\tilde \chi^+} + \Delta M_{s,d}^{\tilde g}
+ \Delta M_{s,d}^{\tilde g \tilde \chi^0} + \Delta M_{s,d}^{\tilde \chi^0}
+ \Delta M_{s,d}^{\rm DP}~,~~~~~~~
\eeqn
\ie box diagrams with the SM contribution, with charged Higgses, charginos, gluinos,
gluino-neutralino and neutralinos and finally neutral Higgs double-penguins, respectively. 
Explicit formulae for $\D M_{s,d}$ that include all the important contributions are given in
\cite{BCRSbig}.\footnote{We mention that gluino and neutralino box contributions 
were not considered in \cite{BCRSbig}. In fact, we find these contributions to be very small, 
but still included them in our numerical analysis. We did not include, instead, subleading effects 
in the Higgs propagator appearing in DP diagrams. The latter have been recently addressed in 
\cite{FGH} and play an insignificant role in the present analysis.}
The values of $B_{s,d}^i$ are taken from \cite{Bparams-lattice-DB}. Besides the dominant SM 
contribution, the most important NP contributions in the DR model are 
$\Delta M_{s,d}^{\tilde \chi^+}$ and especially $\Delta M_{s,d}^{\rm DP}$. The latter is 
strictly negative \cite{BCRSbig}
\beqn
\Delta M_s^{\rm DP}=-12.0 \, {\rm ps}^{-1} 
  \Bigg[\frac{\tan\beta}{50}\Bigg]^4
  \Bigg[\frac{F_{B_s}}{230\,\rm MeV}\Bigg]^2 
  \Bigg[\frac{|V_{ts}|}{0.040}\Bigg]^2\phantom{aaaaaa}\nonumber\\
\times
\Bigg[\frac{\overline m_b(\mu_t)}{3.0 \, {\rm GeV}}\Bigg]
\Bigg[\frac{\overline m_s(\mu_t)}{0.06 \, {\rm GeV}}\Bigg]
\Bigg[\frac{\overline m_t^4(\mu_t)}{M_W^2 \, M^2_A}\Bigg] 
\frac{(16\pi^2 \epsilon_Y)^2}{(1+\tilde\epsilon_3\tan\beta)^2(1+\epsilon_0\tan\beta)^2}~.
\eeqn
We recall that, experimentally \cite{DeltaMs-exp},
\begin{equation*}
(\Delta M_s)_{\rm exp}=(17.77\pm0.10\pm0.07)/ {\rm ps}~,
\end{equation*}
to be compared with the UTfit and CKMfitter SM predictions \cite{UTfit,CKMfitter}
\beqn
(\Delta M_s)^{\rm SM}_{\rm UTfit}=(18.6\pm2.3)/{\rm ps}~,
\qquad
(\Delta M_s)^{\rm SM}_{\rm CKMfitter}=(18.9^{+5.9}_{-2.8})/ {\rm ps}~.
\eeqn

The CKMfitter result still allows for very sizeable NP contributions, while the UTfit 
result bounds $\vert \Delta M_s\vert ^{\rm DP}$ to be below $\approx 3/$ps provided other NP 
contributions in (\ref{DMsformula}) can be neglected. We will see in Section \ref{sec:fit} 
that in the DR model $\Delta M_s$ is slightly suppressed relative to the SM, but this 
suppression amounts to at most 5\%, in accordance with experimental findings.

Concerning $\Delta M_d^{\rm DP}$, it is suppressed by at least two orders of magnitude 
relative to $\Delta M_s^{\rm DP}$ due to $m_d/m_s$ and 
$\vert V_{td} \vert^{2}/\vert V_{ts} \vert^{2}$ factors and consequently in the DR 
model $\Delta M_d$ is SM-like to a very good accuracy.

Finally, we note the role played in the present model by the strong correlation \cite{BCRS-PL} 
between the Higgs penguin contributions to ${\rm BR}(B_{s,d}\to\mu^+\mu^-)$ and $\Delta M_s$: 
the enhancement of ${\rm BR}(B_{s,d}\to\mu^+\mu^-)$ in the DR model is correlated with a 
suppression of $\Delta M_s$. However, the data on $\Delta M_s$ do not allow for very large 
enhancements of ${\rm BR}(B_{s,d}\to\mu^+\mu^-)$ so that both observables turn into
constraints on the pseudoscalar Higgs mass $M_A$. We will come back to this point in
Section \ref{sec:picture}.

\subsection[$B\to X_s \gamma$]{\boldmath $B\to X_s \gamma$} \label{sec:bsg}

An important constraint on any NP model is the inclusive decay $B\to X_s \gamma$
for which the data read \cite{Belle-bsgamma,Babar-bsgamma,HFAG-bsgamma}
\beqn \label{bsg-exp}
{\rm BR}(B\to X_s \gamma)_{\rm exp}=(3.55\pm 0.24\pm 0.10\pm 0.03)\times 10^{-4}~,
\eeqn
to be compared with the SM value at the NNLO level \cite{Misiak-NNLO}
\beqn
{\rm BR}(B\to X_s \gamma)_{\rm SM}=(3.15\pm 0.23)\times 10^{-4}~.
\eeqn

The inclusion of certain non-perturbative effects decreases this value to 
$(2.98 \pm 0.26)\times 10^{-4}$ \cite{becher-neubert}. The latter value, if confirmed, would put some 
tension on the SM prediction. In any case, unless the central experimental value in (\ref{bsg-exp}) 
will be significantly decreased, NP scenarios predicting ${\rm BR} (B\to X_s \gamma)$ 
to be smaller than the SM value are disfavoured.

Instead of presenting detailed formulae for $B\to X_s \gamma$ in the DR model, which can 
be found in the literature \cite{BMU,bsgammaMSSM1,bsgammaMSSM2,MFV}, we collect here a number of qualitative 
properties of these formulae that will turn out to be useful in understanding our numerical results. 
These properties are as follows:
\begin{itemize}
\item  The charged and neutral Higgs contributions to ${\rm BR} (B\to X_s \gamma)$ are 
strictly positive.
\item The sign of the chargino contributions relative to the SM is ruled by the following
relation
\beqn
C_7^{\tilde \chi^+} \propto + \mu A_t \tan\beta \times {\rm sign}(C_7^{\rm SM})~,
\label{C7chi}
\eeqn
with a positive proportionality factor, so it is opposite to that of the SM one
for $\mu > 0$ and $A_t < 0$ (cf. footnote~\ref{fn:A-sign}).
\end{itemize}
This shows that the large $\tan\beta$ effects in $B \to X_s \gamma$ are not as strong as 
in $B_s \rightarrow \mu^{+} \mu^{-}$, where the amplitude behaves as $\tan^3\beta /M_A^2$.
However, they are typically more important than in $\Delta M_s$, since in the latter case 
contributions of Higgs penguins, while behaving as $\tan^4\beta /M_A^2$, are suppressed by
the ratio of the external quark masses over $M_W^2$.

Among the NP contributions to $B \to X_s \gamma$, those from charginos are generically the
largest. In fact, the lightest chargino mass is roughly set by the lowest between 
$\mu$ and $M_2$ and in the DR model it turns out to be generically below $\approx$ 200 GeV.
On the other hand, Higgs contributions are generically small in the DR model, since 
$M_{A,H^+}$ are pushed up by the $B_{s}\rightarrow \mu^{+} \mu^{-}$ constraint.
Consequently, for positive $\mu$, the sign of the Wilson coefficient $C_7$ can be reversed 
relative to $C_7^{\rm SM}$, while $\Delta M_s$, as stated above, cannot be modified by more 
than $\approx 5$\% if the constraint from $B_{s}\rightarrow \mu^{+} \mu^{-}$ is taken into 
account.

Indeed, for positive $\mu$, in order to pass the $B\to X_s \gamma$ constraint, the model favours 
the solution $C_7(\mu_b) = - C_7^{\rm SM}(\mu_b)$. We stress here that such solution is a highly 
conspired one, since such equality should hold at the $\mu_b$ scale, \ie after
running of the coefficients from the matching scale. In addition, in this case SUSY is not
quite a correction to the SM result, but rather the opposite. As a consequence, to address
this case, one would need a theoretical control on the SUSY part at least as good as that
on the pure SM calculation. This task is in turn very hard to achieve, since \egv one
would have to accurately know where to integrate out the various sectors of the MSSM
entering the SUSY contributions to $B\to X_s \gamma$. In absence of such knowledge, one
can take the approach of matching the whole SUSY spectrum at a common reasonable scale.
This approach works if SUSY is a correction to the SM. But in the present case, the
solution $C_7(\mu_b) = - C_7^{\rm SM}(\mu_b)$ is {\em extremely} sensitive to variation 
on the matching scale, and the theoretical error associated completely out of control. We
will come back to this point in section~\ref{sec:picture}.

\subsection[$B\to X_s \ell^{+} \ell^{-}$]{\boldmath $B\to X_s \ell^{+} \ell^{-}$}
\label{sec:Bsll}

An important observable in our analysis will be the branching ratio for the inclusive decay 
$B\to X_s \ell^{+} \ell^{-}$ and the related forward-backward asymmetry $A_{\rm FB}$.
A significant progress in calculating this decay and its exclusive counterpart 
$B\to K^* \ell^{+} \ell^{-}$ has been achieved in recent years through the calculation 
of the NNLO QCD corrections.
The corresponding formulae are very complicated and will not be presented here. They can 
be found in
\cite{BMU-NNLO-bsll,Asatryan-bsll1,Asatryan-bsll2,GHIY,GGH,BGGH,HLMW-NNLO-bsll,Beneke1,Beneke2}. 
For our discussion it will be sufficient to recall the NLO formulae \cite{Misiak-bsll-SM,BurasMunz-bsll-SM}, 
keeping only the dipole operator and the operators
\beqn
Q_{9}=(\ov s b)_{V-A}(\ov \mu \mu)_{V}~, ~~~~ Q_{10}=(\ov s b)_{V-A}(\ov \mu \mu)_{A}~.
\label{Q9-10}
\eeqn
The contributions of semi-leptonic scalar operators to this decay are much less important than in 
$B_{d,s}\rightarrow \mu^{+} \mu^{-}$, since $B\to X_s \ell^{+} \ell^{-}$ is not helicity 
suppressed. We have taken them into account following \cite{hiller-krueger}, but their 
inclusion in the present discussion would only complicate matters without changing 
the basic picture.

Introducing the normalized dilepton mass parameter
\beqn
\hat s = \frac{(p_{\mu^+} + p_{\mu^-})^2}{m_b^2}\equiv \frac{s}{m_b^2}~,
\eeqn
the invariant dilepton mass spectrum in the inclusive decay $B\rightarrow X_s \ell^{+} \ell^{-}$ 
is roughly given at NLO as follows
\beqn
\frac{d \Gamma (B\rightarrow X_s \ell^{+} \ell^{-})}{d \hat s} 
\sim (1- \hat s)^2 \left|V_{ts}\right|^2 U(\hat s)~,
\eeqn
where 
\beqn \label{Us}
U(\hat s)=
(1+2\hat s)\left(|\tilde C_9^{\rm eff}(\hat s)|^2 + |\tilde C_{10}|^2\right) + 
4 \left( 1 + \frac{2}{\hat s}\right) |C_{7}^{(0){\rm eff}}|^2 + 12  \,
C_{7}^{(0){\rm eff}} \, {\rm Re}(\tilde C_9^{\rm eff}(\hat s))~,~~~~~
\eeqn
with the MSSM expression for the Wilson coefficients $C_{7}$ \cite{BMU} and the SM ones for
$C_9, C_{10}$ given in \cite{Misiak-bsll-SM,BurasMunz-bsll-SM,BBL}\footnote{SUSY contributions to $C_9$ and $C_{10}$
\cite{BBE-NNLO,ALGH} are completely negligible in the DR model.}.

Of particular interest is the (normalized) forward-backward asymmetry in $B\rightarrow X_s \ell^{+} \ell^{-}$. 
It becomes non-zero only at the NLO level. It is given in this approximation as follows
\cite{GHIY}
\beqn
\ov A_{\rm FB}(\hat s)= -3\,{\rm Re}\,\left[ \tilde C^*_{10}
\frac{\hat s \, \tilde C_9^{\rm eff}(\hat s)
+2 C_{7}^{(0){\rm eff}}}{U(\hat s)}\right]~.
\eeqn

The expression for the corresponding asymmetry in the exclusive decay $B\to K^* \ell^{+} \ell^{-}$ 
can be found in \cite{Beneke1,Beneke2}. Both asymmetries vanish in the SM at a certain $\hat s=\hat s_0$ \cite{Burdman}, 
which in the case of the inclusive decay is determined through
\beqn
\hat s_0 \, {\rm Re} \, \tilde C_9^{\rm eff}(\hat s_0)+2 C_{7}^{(0){\rm eff}}=0~.
\eeqn

In the SM at NLO one finds $\hat s_0 \approx 0.14$. At NNLO the corresponding value is 
$\hat s_0 = 0.162\pm0.008$
\cite{Asatryan-bsll1,Asatryan-bsll2,Asatrian-bsll3,Asatrian-bsll4,GHIY}.

Now, as can be seen in (\ref{Us}), the very low-$s$ region ($s<1 \,{\rm GeV}^{2}$) is dominated by the 
coefficient $C_{7}$ and does not provide more information than already contained in the 
$B\to X_s \gamma$ decay.
Much more useful is then the low-$s$ region $(1 \,{\rm GeV}^{2}< s < 6 \,{\rm GeV}^{2})$ which 
is theoretically cleaner than the high-s region, is dominated by the Wilson coefficients $C_9$ 
and $C_{10}$ and is also very sensitive to the $C_7$-$C_9$ interference in (\ref{Us}). For 
this low-$s$ range the world average coming from Belle \cite{Belle-bsll} and BaBar
\cite{Babar-bsll} reads
\beqn
{\rm BR}(B\to X_s \ell^{+} \ell^{-})_{\rm exp}=(1.60\pm0.51)\times 10^{-6}~.
\label{bsll-exp}
\eeqn

Concerning the forward-backward asymmetry $A_{\rm FB}$, the only existing positive data come from Belle
\cite{Belle-AFB}, not yet precise enough to be conclusive on the presence of the zero.

For our forthcoming discussion, it will be useful to collect the following general 
properties of $B\to X_s \ell^{+} \ell^{-}$ in the DR model that will be refined later on:
\begin{itemize}
\item  The Wilson coefficients $C_9$ and $C_{10}$ receive only small NP contributions so that the departures of 
$B\to X_s \ell^{+} \ell^{-}$, $A_{\rm FB}$ and $\hat s_0$ from their SM values are governed 
by the modifications of $C_7$.
\item  Recalling that $C_9^{\rm SM}$ and $C_7^{\rm SM}$ have opposite sign, we observe that the flip of 
the sign of $C_7$ by NP contributions will strongly enhance BR$(B\to X_s \ell^{+} \ell^{-})$, 
while the enhancement of $\left \vert C_7 \right \vert$ without the flip of its sign will suppress 
this branching ratio relative to the SM value \cite{HLMW-NNLO-bsll,LPV}
\beqn
{\rm BR}(B\to X_s \ell^{+} \ell^{-})_{{\rm SM}}=(1.58\pm 0.10)\times 10^{-6}~,
\eeqn
that is in perfect agreement with experiment.
\item The value of $\hat s_0$ is correlated with BR$(B\to X_s \gamma)$ if $C_7$ has the SM sign. It increases 
with increasing BR$(B\to X_s \gamma)$.
This is a direct consequence of small NP contributions to $C_9$ in most NP models as pointed 
out in \cite{BPSW}.
\item There is no zero in $A_{\rm FB}$ for ${\rm sign}(C_7)=-{\rm sign}(C_7^{\rm SM})$.

\end{itemize}

\subsection[$B^+ \to \tau^+ \nu$]{\boldmath $B^+ \to \tau^+ \nu$}

Finally, we consider the tree-level decay $B^+ \to \tau^+ \nu$. In the SM, its branching
ratio is simply given as follows,
\beqn
{\rm BR}(B^+ \to \tau^+ \nu)_{\rm SM} = \frac{G_F^2 m_{B^+} M_{\tau}^2}{8\pi}
\left(1-\frac{M_{\tau}^2}{m^2_{B^+}} \right)^2 F_{B^+}^2 |V_{ub}|^2 \tau_{B^+}~.
\eeqn

As the decay constant $F_{B^+}\approx F_{B_d}$ has still sizeable uncertainties, we
consider instead the following ratios \cite{Ikado},
\beqn 
\label{btnu/DMd}
&&\frac{{\rm BR}(B^+ \to \tau^+ \nu)_{{\rm SM}}}{\tau_{B^+}(\Delta M_d)_{{\rm SM}}} = 
\frac{3 \pi}{4 \, \eta_B \, S_0(\ov{m}_t) \, \hat B_{B_d}} \frac{M_\tau^{2}}{M_W^2} 
\left(1 - \frac{M_{\tau}^{2}}{m_{B^+}^2} \right)^2 \left \vert \frac{V_{ub}}{V_{td}}
\right \vert ^2~, \\
\label{btnu/DMs}
&&\frac{{\rm BR}(B^+ \to \tau^+ \nu)_{{\rm SM}}}{\tau_{B^+}(\Delta M_s)_{{\rm SM}}} = 
\frac{3 \pi}{4 \, \eta_B \, S_0(\ov{m}_t) \, \hat B_{B_d}} \frac{M_\tau^{2}}{M_W^2} 
\frac{1}{\xi^2} \frac{m_{B^{+}}}{m_{B_{s}}}
\left(1 - \frac{M_{\tau}^{2}}{m_{B^{+}}^2} \right)^2 \left \vert \frac{V_{ub}}{V_{cb}} 
\right \vert^2~,~~~~~~
\eeqn
where we used $F_{B_d}\approx F_{B^+}$, $m_{B_d}\approx m_{B^+}$, 
$\vert V_{ts} \vert  \approx \vert V_{cb} \vert$ and the ratio $\xi$ defined as
\beqn
\xi = \frac{F_{B_s}\sqrt{\hat B_{B_s}}}{F_{B_d}\sqrt{\hat B_{B_d}}} ~ .
\eeqn

The uncertainties on the right-hand side of (\ref{btnu/DMd}) and (\ref{btnu/DMs}) are comparable. In 
(\ref{btnu/DMd}), the only hadronic uncertainty resides in $\hat B_{B_d}$, while in (\ref{btnu/DMs}) 
there is an additional uncertainty in $\xi$.
On the other hand, the ratio $\left \vert V_{ub}/V_{cb} \right \vert ^2$ can be determined 
from tree-level decays without NP pollution, while $\left \vert V_{td} \right \vert ^2$ can 
clearly be affected by NP contributions.

Using the input parameters of Table \ref{tab:btaunu} and \ref{tab:obs}, we find from (\ref{btnu/DMd})
\beqn
{\rm BR}(B^+ \to \tau^+ \nu)_{\rm SM}= \left\lbrace 
\begin{array}{lll} 
(0.87\pm0.11)\times10^{-4}~, ~~~~\left \vert V_{ub} \right \vert_{\rm UTfit}~,\\ 
(1.31\pm0.23)\times10^{-4}~, ~~~~\left \vert V_{ub} \right \vert _{\rm incl}~, \\
\end{array} \right.
\eeqn
where the first estimate uses the value for $\left \vert V_{ub} \right \vert $ resulting from 
the UTfit analysis of \cite{UTfit}, while 
$|V_{ub}|_{\rm incl}$ is the value resulting from inclusive decays alone. 
The corresponding values from (\ref{btnu/DMs}) read
\beqn
{\rm BR}(B^+ \to \tau^+ \nu)_{\rm SM}= 
\left\lbrace \begin{array}{lll} (0.82\pm0.12)\times 10^{-4}&~,~~~~\left \vert V_{ub} \right \vert_{\rm UTfit}~,\\
(1.24\pm0.24)\times10^{-4}&~,~~~~\left \vert V_{ub} \right \vert_{\rm incl}~, \\
\end{array} \right.
\eeqn
showing that the formulae (\ref{btnu/DMd}) and (\ref{btnu/DMs}) give similar results.

\TABLE[t]{
\begin{tabular}{|lcr||lcr|}
\hline
Parameter & Value & Ref. & Parameter & Value & Ref. \\
\hline \hline
$10^{3}|V_{ub}|_{\rm UTfit}$ & $3.66(15)$ & \cite{UTfit} & $\Delta M_s [{\rm ps^{-1}}]$ & $17.77(12)$ & \cite{HFAG}\\
$10^{3}|V_{ub}|_{\rm incl}$ & $4.49(33)$ & \cite{HFAG} & $\Delta M_d [{\rm ps^{-1}}]$ & $0.507(5)$ & \cite{HFAG}\\
$10^{3}|V_{td}|$ & $8.49(28)$ & \cite{UTfit} & $\ov{m}_t$ & $161.7(2.0)$ & \\
$\hat B_{B_d}$ & $1.28(9)$ &  & $m_{B_s}$ & $5.3661(6)$ & \cite{PDBook}\\
$\xi$ & $1.23(6)$ & \cite{lattice} & $m_{B^+}$ & $5.27913(31)$ & \cite{PDBook}\\
$\eta_B$ & $0.55$ &  & $\tau_{B^+} [{\rm 10^{-12}s}]$ & $1.638(11)$ & \cite{PDBook}\\
\hline \hline
\end{tabular}
\caption{Input parameters for the SM prediction of ${\rm BR}(B^+ \to \tau^+ \nu)$. Dimensionful quantities are expressed in GeV, unless otherwise specified.}
\label{tab:btaunu}}

We observe that the theoretical branching ratio with $\left \vert V_{ub} \right \vert
_{\rm incl}$ is closer to the experimental average between the Belle \cite{Belle-btaunu} 
and BaBar \cite{Babar-btaunu} results, which reads \cite{UTfit-btaunu}
\beqn
{\rm BR}(B^+ \to \tau^+ \nu)_{\rm exp}=(1.31\pm0.48)\times10^{-4}~,
\label{btaunu-exp}
\eeqn
but the large experimental error precludes any clear cut conclusions at present. Yet,
similarly to the case of the $B\to X_s \gamma$ decay, extensions of the SM that predict 
${\rm BR}(B^+ \to \tau^+ \nu)$ to be smaller than the SM value seem to be disfavoured 
at present.

In this respect two-Higgs-doublet models of type-II, like the MSSM, where each doublet 
couples separately to up- and down-type quarks, are interesting as the interference 
between $W$ and $H^+$ amplitudes is necessarily destructive \cite{Hou}.
One finds then \cite{Ake-Reck,isidori-paradisi}
\beqn
R_{B\tau \nu} = \frac{{\rm BR}(B^+ \to \tau^+ \nu)^{\rm DR}}{{\rm BR}(
B^+ \to \tau^+ \nu)^{\rm SM}} = \left[1 - \frac{m_{B^{+}}^2}{m_{H^{+}}^2}  
\frac{\tan^2 \beta}{1+\epsilon_0 \tan \beta}\right]^2 \left \vert \frac{V_{ub}^{\rm
DR}}{V_{ub}^{\rm SM}} \right \vert ^2~. 
\eeqn
We have explicitly shown the dependence on $V_{ub}$ since, in the DR model, the value
of $\left \vert V_{ub} \right \vert$ turns out to be even smaller than 
$\left \vert V_{ub} \right \vert$ extracted from the UT SM fit.

\subsection[$(g-2)_{\mu}$]{\boldmath $(g-2)_{\mu}$}

In principle we should also consider $(g-2)_{\mu}$, where the data seem to be above 
the SM expectations by roughly $3\sigma$. In many supersymmetric models one finds, for large $\tan\beta$, $\mu>0$ 
and slepton masses O(400 GeV), additional contributions to $(g-2)_{\mu}$ 
that allow to fit the data. However, in the DR model, the slepton masses are larger than 
1 TeV and the NP contribution amounts to at most one $\sigma$ of the SM value.
Therefore the DR model cannot fit the present data on $(g-2)_{\mu}$ and we will not 
include this observable in the global fit, keeping also in mind that the theoretical status 
of $(g-2)_{\mu}$ is not yet fully satisfactory \cite{g-2muon,g-2muon2,g-2muon3}.

\section{General picture} \label{sec:picture}

Having the formulae for the FC observables at hand, we can discuss first the general
pattern of these observables within the DR model. A detailed numerical analysis will
be presented in the next section.

\subsubsection*{Step 1}

In the DR model, due to the unification of Yukawa couplings, $\tan\beta$ is 
forced to be close to 50. This fact, as already stressed in section \ref{sec:FCNC}, requires 
$M_A$ to be sufficiently large in order for the predicted branching ratio of $B_s\to\mu^+\mu^-$ 
in (\ref{bsmumuSUSY}) to be consistent with the upper bound in (\ref{bsmumu-exp}).
Typically we find $M_A > 450~\text{GeV}$. For such large Higgs masses, 
one has $M_{H^+} \approx M_A \approx M_{H_0}$ and this bound
is also approximately valid for $M_{H^+}$ and $M_{H_0}$.

\subsubsection*{Step 2}

As already stressed in section \ref{sec:bsg}, the large values of $M_{H^+}$ and $M_A$ imply 
that NP contributions to BR($B\to X_s\gamma$) are dominated by charginos, with the positive charged 
and neutral Higgs contributions (as well as those from gluinos and neutralinos) being subleading. 

With $\mu > 0$ it is possible to fit the data on BR($B\to X_s\gamma$) by making the 
chargino contribution so large that $C_7^{\rm SUSY} \simeq -2 \,C_7^{\rm SM}$ at the
$\mu_b$ scale. There are several problems with this choice. 

First, there is the problem already stressed at the end of section \ref{sec:bsg}.
The possibility of having $C_7^{\rm SUSY} \simeq -2 \,C_7^{\rm SM}$ at the
$\mu_b$ scale implies that the SUSY contribution is not quite a correction to the SM matching
condition, but rather the opposite. With such a large correction coming from NP, the usual 
argument of neglecting NLO QCD corrections to NP contributions becomes invalid. One
would need a control on the NP side {\em at least} as good as the one present in the SM
contribution. This holds not only for the anomalous dimension matrix \cite{CHM}, but also for the
matching conditions of the SUSY contributions \cite{CDGG,BMU,BGY,DGS}. Indeed, with the use of only LO matching
conditions for SUSY, we find a large sensitivity of the finely-tuned condition $C_7^{\rm SUSY} \simeq -
2\,C_7^{\rm SM}$ to the choice of the matching scale for the SUSY contributions.
Such sensitivity becomes even stronger on the resulting BR($B\to X_s\gamma$) and this makes a 
meaningful inclusion of the BR($B\to X_s\gamma$) constraint
in the numerical analysis practically impossible.

Second, there is no zero in the forward-backward asymmetry $A_{\rm FB}$, which is however still 
elusive experimentally \cite{Belle-AFB}.

Third and foremost, it has been shown in \cite{GHM} that this solution for $C_7$ is actually excluded by 
the experimental data on ${\rm BR}(B\to X_s \ell^+\ell^-)$, provided the new physics contributions to the 
Wilson coefficients $\tilde C_9^{\rm eff}$ and $\tilde C_{10}$ are small.
In fact, the maximal ranges of such contributions in the MSSM with MFV found 
in \cite{ALGH} are too small to bring the theory prediction for ${\rm BR}(B\to X_s \ell^+\ell^-)$ 
in accordance with the experimental data. These findings have also been confirmed in \cite{LPV}.

In summary, we want to stress that although we cannot meaningfully take into account the case  
$C_7^{\rm SUSY} \simeq -2 \,C_7^{\rm SM}$ in the numerical analysis due to the theoretical 
uncertainties described above, we can exclude this case and impose $\text{sign}(C_7) 
\equiv \text{sign}(C_7^{\rm SM})$ as a constraint in the global fitting procedure, because of the 
model-independent arguments brought forward in the previous paragraph, which are unaffected by such 
uncertainties.

\subsubsection*{Step 3}

If $\mu$ is chosen to be positive and one 
attempts to be consistent with the data on $B\to X_s \ell^+\ell^-$ in (\ref{bsll-exp}) by keeping 
$C_7$ to have the SM sign, as discussed in the previous step, the negative contribution of charginos 
tends to suppress BR($B\to X_s\gamma$) below acceptable values and can only be tamed by raising the 
squark masses until these contributions decouple.

\subsubsection*{Step 4}

We next move to $\mu<0$. In this case, $C_7$ has the same sign as $C_7^{\rm SM}$, but due to 
constructive interference between SM, scalar and chargino contributions, it tends to be too large, unless 
squarks are sufficiently heavy. In this respect, we note that, for every given $m_{16}$, this case is not 
simply a reflection of the corresponding case with $\mu > 0$. As we will see in the numerical section 
below, for negative $\mu$ the lightest squark masses are generically higher ($\gtrsim 2$ TeV) than in the 
corresponding positive $\mu$ case.

\subsubsection*{Step 5}

Let us finally look at BR($B^+\to\tau^+\nu$). We have seen that in order to bring the SM 
value for this branching ratio close to its central experimental value, it was necessary to choose 
the tree-level value for $|V_{ub}| > 4 \times 10^{-3}$. In the DR model, we do not have this 
freedom as $|V_{ub}|$ is in principle a prediction. In practice it is an outcome of the
global fit to the model parameters and such fit definitely prefers a low value for
$|V_{ub}|$, around $3.2 \times 10^{-3}$. With such value and the negative contribution from 
charged Higgses, we find typically
\beqn
{\rm BR}(B^+\to\tau^+\nu) \leq 0.6 \times 10^{-4}~.
\eeqn
While this is not yet excluded, in view of the large experimental error in (\ref{btaunu-exp}), 
also this decay could turn out to be problematic for the DR model if the central experimental 
value will remain above $1.0 \times 10^{-4}$ and the error will decrease by a factor of two.

\subsubsection*{Final remarks}

In summary, we have shown that while it is possible through choice of the parameters to obtain 
the agreement of the DR model with a given single observable discussed above,
simultaneous agreement for all observables is possible at most with very heavy sfermions.
In the next section, we will present a detailed numerical analysis of these findings.

\section{Numerical analysis} \label{sec:fit}

\subsection{Fitting procedure} \label{sec:proc}

We now turn to describe the numerical strategy adopted to test the DR model. As we have
seen in Section \ref{sec:pars}, the model is completely specified in terms of 24
parameters, listed in Table \ref{tab:parameters} and here collectively indicated as
$\vec\vartheta$. After fixing them, it is possible to reconstruct the {\em whole} MSSM 
at low-energy scales, with a well-controlled theoretical error. The procedure, based on
RGEs, which one adopts to connect the GUT scale model to low energies, has been described in 
Sections \ref{sec:RGE} and \ref{sec:obs}. 

Once the low-energy MSSM is specified, the model is testable. To this end, one needs a
suitable set of observables, whose experimental determinations $\mc O_i$ should be as
precise as possible and, on the theoretical side, calculable within the MSSM with
sufficient accuracy. Since the MSSM is the low-energy result of the GUT scale model, the 
theoretical prediction for the observable $\mc O_i$ will be functions $f_i[\vec\vartheta]$ 
of the model parameters. In order to compare theory predictions with experimental values, 
one defines a suitable $\chi^2$-function as
\beqn
\chi^2[\vec\vartheta] \equiv \sum_{i = 1}^{N_{\rm obs}}
\left( \frac{f_i[\vec\vartheta] -\mc O_i}{\sigma_i} \right)^2~,
\label{chi2}
\eeqn
where the uncertainty $\sigma_i$ associated with the $i^{\rm th}$ observable is
defined as
\beqn
\sigma_i = \sqrt{(\sigma_i^2)_{\rm exp} + (\sigma_i^2)_{\rm theo}}~.
\label{sigma}
\eeqn
Here $(\sigma_i)_{\rm exp}$ is the experimental RMS error and $(\sigma_i)_{\rm theo}$ an
estimate of the theoretical error associated with the $f_i[\vec\vartheta]$ calculation.

The $\chi^2$-function (\ref{chi2}) is then minimized upon variation of the model
parameters $\vec\vartheta$. To this end, we have adopted the minimization algorithm 
{\tt MIGRAD}, which is part of the {\tt CERNlib} library \cite{CERNlib}. The minimum value
for the $\chi^2$-function provides then a quantitative test of the performance of the
model in reproducing the observables entering the fit. We mention here that, strictly
speaking, such test cannot be attached a statistically rigorous meaning, \ie it is not a
`Pearson's test', since, \egv the $\chi^2$-entries are not all independently measured
observables. Nonetheless, the numerical value of the minimum for the function
(\ref{chi2}), as well as the single pulls in its entries, provide a good quantitative
indication of the detailed performance of the model for the single observables.
\TABLE[]{
\begin{tabular}{|lcr||lcr|}
\hline
Observable & Value($\sigma_{\rm exp}$) & Ref. & Observable & Value($\sigma_{\rm exp}$) & Ref. \\
\hline \hline
$M_W$ & $80.403(29)$ & \cite{PDBook} & $M_\tau$ & $1.777(0)$ & \cite{PDBook} \\
$M_Z$ & $91.1876(21)$ & \cite{PDBook} & $M_\mu$ & $0.10566(0)$ & \cite{PDBook} \\
$10^{5} G_\mu$ & $1.16637(1)$ & \cite{PDBook} & $10^3 M_e$ & $0.511(0)$ & \cite{PDBook} \\
$1/\al_\text{em}$ & $137.036$ & \cite{PDBook} & $|V_{us}|$ & $0.2258(14)$ & \cite{UTfit} \\
$\al_s(M_Z)$ & $0.1176(20)$ & \cite{PDBook} & $10^3 |V_{ub}|$ & $4.1(0.4)$ &
\cite{CKMfitter} \\
$M_t$ & $170.9(1.8)$ & \cite{PDBook} & $10^2 |V_{cb}|$ & $4.16(7)$ & \cite{UTfit} \\
$m_b(m_b)$ & $4.20(7)$ & \cite{PDBook} & $\sin 2 \be$ & $0.675(26)$ & \cite{HFAG} \\
$m_c(m_c)$ & $1.25(9)$ & \cite{PDBook} & $10^3 \D m_{31}^2$ [eV$^2$]& $2.6(0.2)$ &
\cite{GonzalezMaltoni} \\
$m_s(2~{\rm GeV})$ & $0.095(25)$ & \cite{PDBook} & $10^5 \D m_{21}^2$ [eV$^2$]& $7.90(0.28)$ & \cite{GonzalezMaltoni} \\
$m_d(2~{\rm GeV})$ & $0.005(2)$ & \cite{PDBook} & $\sin^2 2 \theta_{12}$ & $0.852(32)$ & \cite{GonzalezMaltoni} \\
$m_u(2~{\rm GeV})$ & $0.00225(75)$ & \cite{PDBook} & $\sin^2 2 \theta_{23}$ & $0.996(18)$ & \cite{GonzalezMaltoni} \\
\hline \hline
\end{tabular}
\caption{Flavour conserving observables used in the fit. Dimensionful quantities are expressed in GeV, 
unless otherwise specified.}
\label{tab:obs}
}

The observables used in the fit are reported in Tables \ref{tab:obs}-\ref{tab:bounds}.
Concerning the latter, the following comments are in order.
\begin{itemize}

\item The observables in Table \ref{tab:obs}
were already used -- among the others -- in the previous studies \cite{DR05,DR06} of the DR model,
where the very good performance of the model in fitting them was demonstrated.
We mention that the experimental determination of the observables themselves should 
{\em not} rely on any theoretical assumption which would be invalidated in the presence 
of NP, \ie one should choose observables whose determination is NP-independent. This
comment applies in particular to CKM-related quantities, among which one keeps only those
measured through tree-level processes and $\sin 2 \beta_{\psi K_S}$, which gives direct
access to $\sin 2 \beta$ since the DR model has only one CP phase in the quark sector.

\item The observables in Table \ref{tab:FCobs}, on the other hand, represent the real
novelty of our study with respect to the previous ones. Such FC processes are not calculated
after the fitting procedure, but instead introduced {\em directly} in the
$\chi^2$-function. The procedure to calculate these FC observables has been detailed in
Section \ref{sec:FCNC}.

\item In addition, we included in the fitting function a number of constraints, \ie those 
on the lightest Higgs mass and on the lightest components of the SUSY spectrum, Table
\ref{tab:bounds}, and the constraint on the BR$(B_s \to \mu^+ \mu^-)$, Table \ref{tab:FCobs}. 
These constraints are in the form of suitably smoothened step functions, which are added to 
the $\chi^2$-function of eq. (\ref{chi2}). If any of the constraints is violated, the step
functions add a large positive number to the $\chi^2$, while for respected constraints the
returned value is zero, so that the $\chi^2$ is set back to its `unbiased' definition
(\ref{chi2}).
\end{itemize}

\TABLE[ht]{
\begin{tabular}{|lcr|}
\hline
Observable & Value($\sigma_{\rm exp}$)($\sigma_{\rm theo}$) & Ref. \\
\hline \hline
$10^3 \eps_K$ & 2.229(10)(252) & \cite{PDBook} \\
$\D M_s / \D M_d$ & 35.0(0.4)(3.6) & \cite{HFAG,UTfit} \\
$10^4$ BR$(B \to X_s \gamma)$ & 3.55(26)(46) & \cite{Misiak-NNLO} \\
$10^6$ BR$(B \to X_s \ell^+ \ell^-)~, ~~q^2_{\ell^+ \ell^-} \in [1,6]$ GeV$^2$ & 1.60(51)(40) & \cite{LPV} \\
$10^4$ BR$(B^+ \to \tau^+ \nu)$ & 1.31(48)(9) & \cite{UTfit-btaunu} \\
BR$(B_s \to \mu^+ \mu^-)$ & $< 1.0 \times 10^{-7}$ & \cite{CDF-bsmumu} \\
\hline \hline
\end{tabular}

\caption{FC observables used in the fit.}
\label{tab:FCobs}
}

\TABLE[r]{
\begin{tabular}{|lcr|}
\hline
Observable & Lower Bound & Ref. \\
\hline \hline
$M_{h_0}$ & $114.4$ GeV & \cite{PDBook} \\
$m_{\tilde t}$ & $60$ GeV & \cite{PDBook} \\
$m_{\tilde\chi^+}$ & $104$ GeV & \cite{PDBook} \\
$m_{\tilde g}$ & $195$ GeV & \cite{PDBook} \\
\hline \hline
\end{tabular}

\caption{Mass bounds used in the fit.}
\label{tab:bounds}
}

Further comments on the determination of the theoretical errors are in order. First, one can note 
that among the observables in Table \ref{tab:obs}, some have a negligible experimental error. 
In this case, we took as overall uncertainly 0.5\% of the experimental value, which we consider a
realistic estimate of the numerical error associated with the calculations\footnote{
Note that this error is more conservative than the 0.1\% used in \cite{DR05,DR06}.
}. Concerning the
theoretical errors on the flavour observables (Table \ref{tab:FCobs}), we note the following: 
the error on $\eps_K$ is basically that on the lattice parameter $\hat B_K$; the error on 
$\D M_s / \D M_d$ keeps into account that on the SM contribution, dominated by $\xi^2$ and
that on the NP contributions, dominated by the scalar $P_L \otimes P_R$ operators; the error on 
BR$(B^+ \to \tau^+ \nu)$, after normalization as in eq. (\ref{btnu/DMd}), is only that on 
$\hat B_d$; the error on BR$(B \to X_s \gamma)$ is taken as twice the total theoretical error 
associated with the SM calculation \cite{Misiak-NNLO}; finally the error on BR$(B \to X_s \ell^+ \ell^-)$ 
is taken as 25\% of the experimental result, and is estimated from the spread of the
theoretical predictions after variations of the scale of matching of the SUSY
contributions.

We next turn to the generic strategy adopted to minimize the $\chi^2$-function with
respect to the parameters $\vec\vartheta$. We note that, among them, $m_{H_{u,d}}$, $\mu$
and $\tan \beta$ are those responsible for EW symmetry breaking, and the $\chi^2$-function
manifests a particularly sensitive dependence on them, especially on $m_{H_{u,d}}$. As a
consequence, such parameters are varied first (keeping the other fixed to initial guesses), 
in order to successfully find an EW symmetry breaking minimum, thereafter varying the rest of 
the parameters. This procedure is schematically described in the flow-chart of Fig. 
\ref{fig:chart}. As a final step, all the parameters are varied simultaneously.

\FIGURE[ht]{
\includegraphics[scale=0.8]{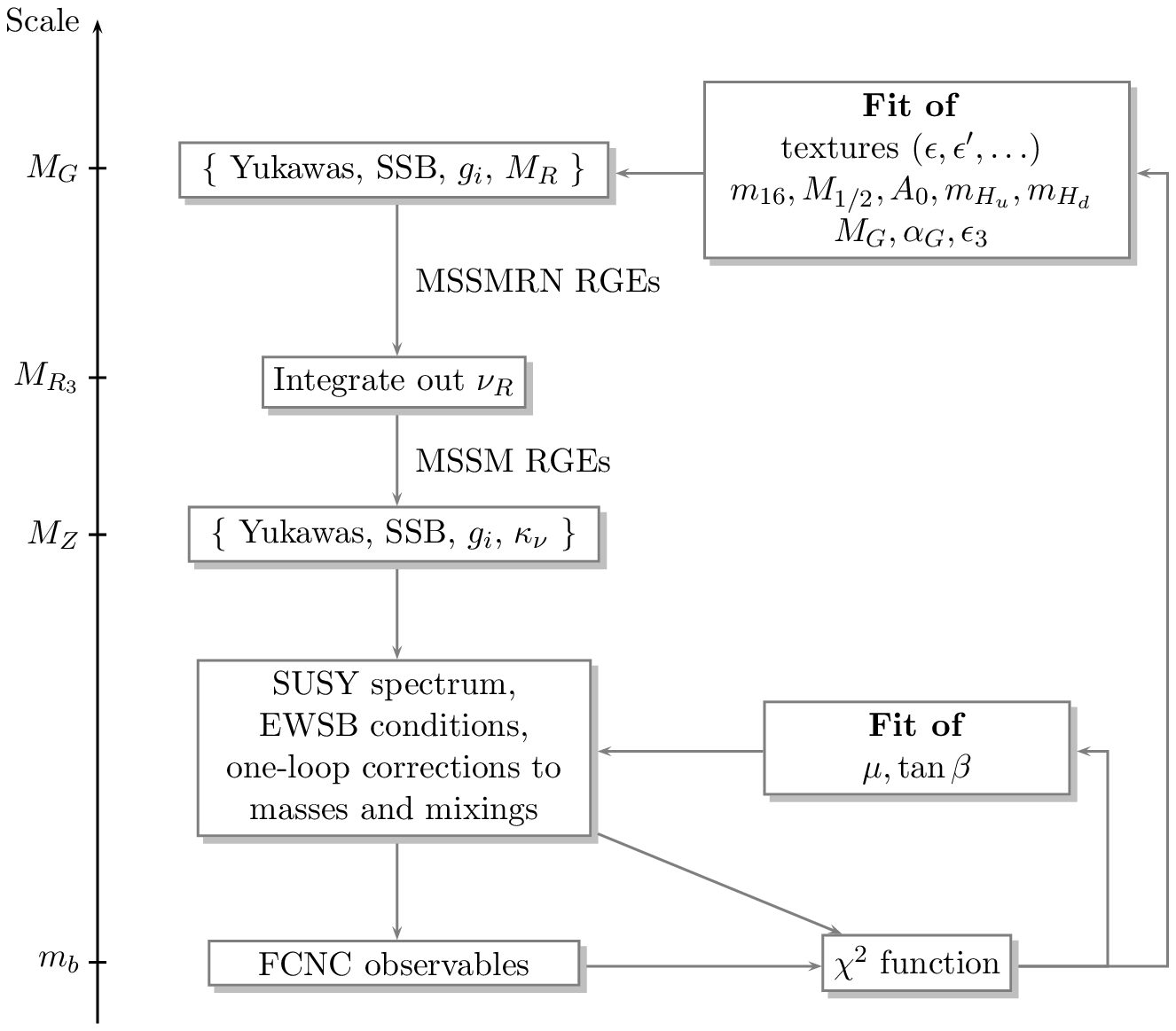}
\caption{Schematic chart of the strategy followed in the fitting procedure.}
\label{fig:chart}
}

Among the model parameters, the ones in the SUSY sector are particularly interesting since they 
set the scale of the SUSY particles' masses. This is especially true for $m_{16}$ and $\mu$.
All the other parameters can be left free in the fit, since their typically allowed range
of variation is quite narrow. In particular, when $m_{16}$ is fixed and $\mu$ is positive, the fit
prefers regions of the remaining parameter space such that
\beqn
\mu, M_{1/2} \ll m_{16}~, ~~~ -A_0 \simeq 2 m_{16}~,
\label{par-constraints}
\eeqn
which is favoured by third generation Yukawa unification \cite{BDR1, BDR2, Auto}. Concerning the first of
relations (\ref{par-constraints}), $M_{1/2}$ is bounded from above because otherwise the bottom mass 
is pushed up beyond acceptable values by large gluino corrections. In fact, we find that $M_{1/2}$ is 
most of the times chosen in the range $[140,400]$ GeV, where the lower bound results from the chargino 
mass bound in Table \ref{tab:bounds}.

The second of relations (\ref{par-constraints}) leads to an inverted scalar mass hierarchy \cite{BFPZ}, 
\ie\ heavy first and second generation sfermions, but lighter third generation sfermions.
For the values of $m_{16}$ considered here, namely $m_{16} \geq 4$ TeV, and for $\mu > 0$,
this condition also helps to obtain the correct prediction for $m_b$ \cite{BDR1,BDR2}.

On the other hand, the allowed interval for $\mu$ is generically wider, for every fixed value 
of $m_{16}$. As a consequence, our main strategy is to study the model behavior for different choices
of $\{m_{16}, \mu\}$, and for each of them, let the rest of the parameter space free to be
determined by the minimization procedure. In the next section we now turn to describe the
various scenarios considered in the $\{m_{16}, \mu\}$ plane.

\subsection{Scenarios} \label{sec:scenarios}

We considered increasing values of $m_{16}$ starting from 4 TeV, which represents the
`minimum' value for successful fits to the observables of Table \ref{tab:obs}
\cite{DR05,DR06}. For each fixed value of $m_{16}$ we then studied the $\mu$ dependence by 
performing fits with different initial guesses for this parameter. All input values for the model 
parameters corresponding to these scenarios are listed in Table~\ref{tab:fit-par}. In the following, we describe 
in detail our findings.

\subsubsection[$m_{16} = 4$ TeV, $\mu > 0$]{\boldmath $m_{16} = 4$ TeV, $\mu > 0$} \label{sec:4000_mu+}

Given the inverted scalar mass hierarchy, the relatively low value of $m_{16}$
leads to stop masses below 1~TeV, resulting in a large chargino contribution 
to BR$(B \to X_s \gamma)$.
\FIGURE{
\includegraphics[scale=0.7]{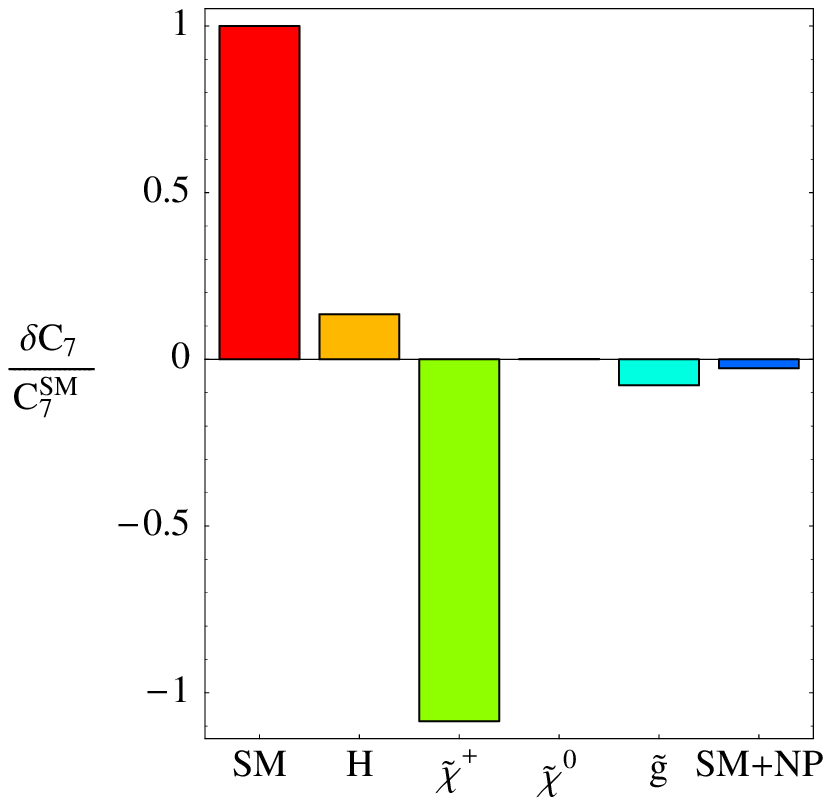}
\caption{Contributions to $C_7(\mu_b)$ for the scenario of section \ref{sec:4000_mu+}.}
\label{fig:bsg_m164000_mu+}
}
As a matter of fact, the preferred region of NP contributions
to the latter decay mode reverses the sign for $C_7(\mu_b)$: $C_7(\mu_b) \simeq - C_7^{\rm SM}(\mu_b)$. 
On the fine-tuned character of this case we have already commented
in Sections \ref{sec:bsg} and \ref{sec:picture}. We stress again that within the DR model, this
solution is not viable in view of the implied enhancement to the branching ratio of
$B \to X_s \ell^+ \ell^-$. Model-independent analyses \cite{GHM,LPV} show in fact that to compensate for
such enhancement, one would need substantial contributions to the Wilson coefficients
$\tilde C_9^{\rm eff}$ and $\tilde C_{10}$ of the operators in eq. (\ref{Q9-10}).
However, within the DR model, these Wilson coefficients are always SM-like to an excellent
approximation.

As a consequence of the above, we have studied the viability of
having the SM sign in $C_7(\mu_b)$
by imposing this condition as a constraint on the $\chi^2$ function. The typical fit result
in this case is illustrated in Table \ref{tab:m164000_mu+}.

The fit displays the main problem of the model in this regime for $\{m_{16},\mu\}$, \ie
a $5\sigma$ discrepancy in the predicted BR$(B \to X_s \gamma)$. Even imposing
the SM sign on $C_7(\mu_b)$,
the contribution from charginos is still too large in magnitude, and none of the other NP contributions 
is able to compensate for it. The situation is illustrated in Fig.~\ref{fig:bsg_m164000_mu+}.

A second, though less severe, problem is the predicted value for BR$(B^+ \to \tau^+ \nu)$,
which is roughly $2\sigma$ too low with respect to the experimental average
(\ref{btaunu-exp}). This problem is strictly connected to the quite low value for
$|V_{ub}| \approx 3.2 \times 10^{-3}$ predicted by the model. We found this feature to hold
irrespective of the values chosen for $\{m_{16},\mu\}$, so that it should be connected
to the specific Yukawa textures of the model.

\subsubsection[$m_{16} = 6$ TeV, $\mu > 0$]{\boldmath $m_{16} = 6$ TeV, $\mu > 0$}
\label{sec:6000_mu+}

Since the BR$(B \to X_s \gamma)$ problem is related to the low value for $\mu$ 
required (at least for positive $\mu$) by $m_{16} = 4$ TeV, we have tried to increase the
latter in order to understand how fastly decoupling is effective in mildening the problem.
\FIGURE[ht]{
\includegraphics[scale=0.7]{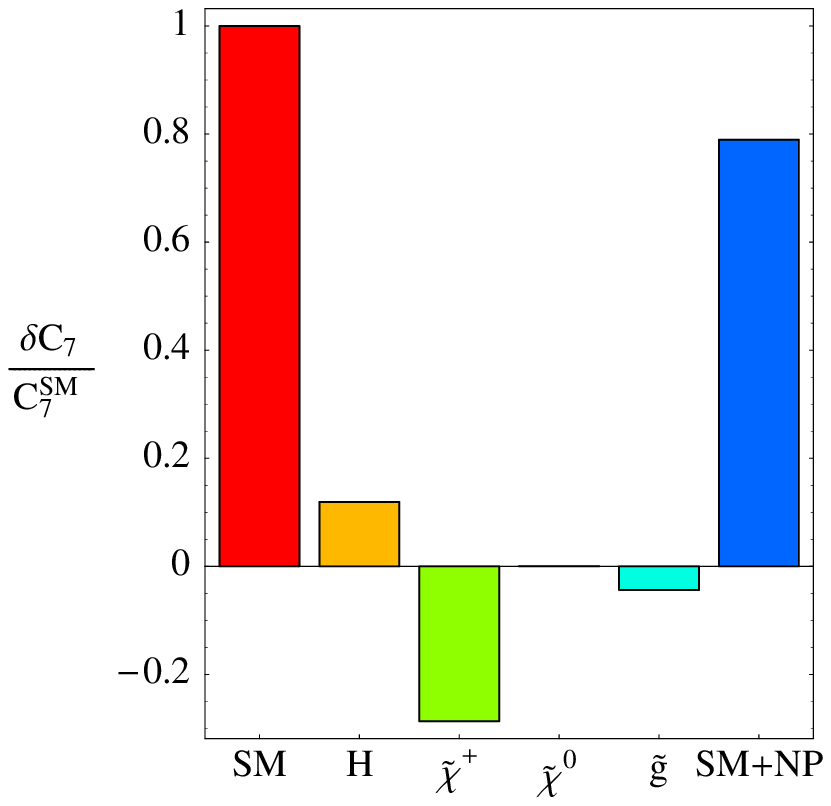}
\caption{Contributions to $C_7(\mu_b)$ for the scenario of section \ref{sec:6000_mu+}.}
\label{fig:bsg_m166000_mu+}
}

For $m_{16} = 6$ TeV and $\mu > 0$ the preferred range for $\mu$ is roughly $\mu \in [800,
1000]$ GeV. A typical fit result is displayed in Table \ref{tab:m166000_mu+},
corresponding to the case $\mu = 953$ GeV. As a matter of fact, the discrepancy in $B \to X_s \gamma$ 
is tamed to roughly $2.3\sigma$, since chargino contributions are less important than
in the $m_{16} = 4$ TeV cases. The various contributions to $C_7(\mu_b)$ for the fit in
Table \ref{tab:m166000_mu+} are displayed in Fig. \ref{fig:bsg_m166000_mu+}. By decreasing
$\mu$ below $\approx 800$ GeV, the prediction for $B \to X_s \gamma$ fastly worsens. For example, 
for a converged fit with $\mu = 430$ GeV, we found that the discrepancy is already at the $4.2\sigma$ level. 
Concerning $B^+ \to \tau^+ \nu$, as anticipated above, the predicted rate remains always roughly $2\sigma$ off.

As a final remark, while the case $m_{16} = 6$ TeV allows, for suitable $\mu$, a smaller discrepancy 
in $B \to X_s \gamma$, the latter comes with the price of a much higher mass for the lightest up-type squark, 
as evident by comparing the corresponding values in Table \ref{tab:predictions}. 

\subsubsection[$m_{16} = 10$ TeV, $\mu > 0$]{\boldmath $m_{16} = 10$ TeV, $\mu > 0$} \label{sec:10000_mu+}

Increasing $m_{16}$ further, the chargino contribution to $\text{BR}(B\to X_s\gamma)$ becomes comparable 
in size to the charged Higgs contribution, resulting in acceptable values for this branching ratio. 
For example, in the fit of Table~\ref{tab:m1610000_mu+}, the pull in this observable is reduced to 
$1.3\sigma$. On the other hand, the pull resulting from $B^+\to\tau^+\nu$ is not
significantly ameliorated compared to the previous cases. In addition, the lightest squark is as heavy 
as 1.9~TeV.

\subsubsection[$m_{16} = 4$ TeV, $\mu < 0$]{\boldmath $m_{16} = 4$ TeV, $\mu < 0$} \label{sec:4000_mu-}

We explored also the case with negative $\mu$. In this instance, relations in eq.
(\ref{par-constraints}) (with $\mu \to |\mu|$) do not apparently need to be fulfilled. As
a matter of fact, we found satisfactory fits for a quite wide range of $\mu$: $\mu
\in -[2100,400]$ and $A_0$ is always lower in magnitude than the value required by the
second relation in eq. (\ref{par-constraints}), typically leading to very small $A_t$. As a consequence, 
the squark mass spectrum does not fulfill an inverted mass hierarchy \cite{BFPZ,Auto} and squark 
masses are generically very heavy. In this case, also heavy Higgses are found to have
generically larger masses, $\gtrsim 1.5$ TeV. A typical result is shown in Table \ref{tab:m164000_mu-} 
and the displayed features remain basically the same in the full allowed range for $\mu$.

We observe that, in this case, negative values of $\mu$, large squark masses and small 
values for $A_t$ imply small threshold corrections to $m_b$ and therefore allow to have
successful Yukawa unification away from the inverted mass hierarchy condition. We have
then investigated whether an acceptable fit away from inverted mass hierarchy could also
be obtained for $\mu > 0$\footnote{We warmly thank R. \dmsk\ for drawing this point to our 
attention.}. In the latter case, the most important contributions to the $m_b$ threshold correction 
have the same sign and consequently one generically needs larger squark masses than for
$\mu < 0$, in order to reproduce the right $m_b$ value. In fact 
we find that, unless $m_{16} \gtrsim 6$ TeV, the prediction on $m_b$ is 4$\sigma$ too 
large\footnote{One should also take into account the quite precisely known value for $m_b$ 
assumed in the present paper.} and consequently fits with $\mu>0$ and away from the inverted mass 
hierarchy perform worse than the corresponding negative $\mu$ cases.

We also note that, in the $\{m_{16},\mu\}$ mass scenario considered in the present subsection, the predicted 
value for BR$(B \to X_s \gamma)$ is always larger than the SM prediction and close to the experimental 
value. In fact, small $A_t$ implies negligible chargino contributions, so that the
main correction is the one from Higgses. However, since the lightest stop is around 2.6 TeV, this scenario 
clashes with the motivation for SUSY as a solution to the Higgs fine-tuning problem.

\subsection{Results}

\FIGURE{
\includegraphics[width=7cm]{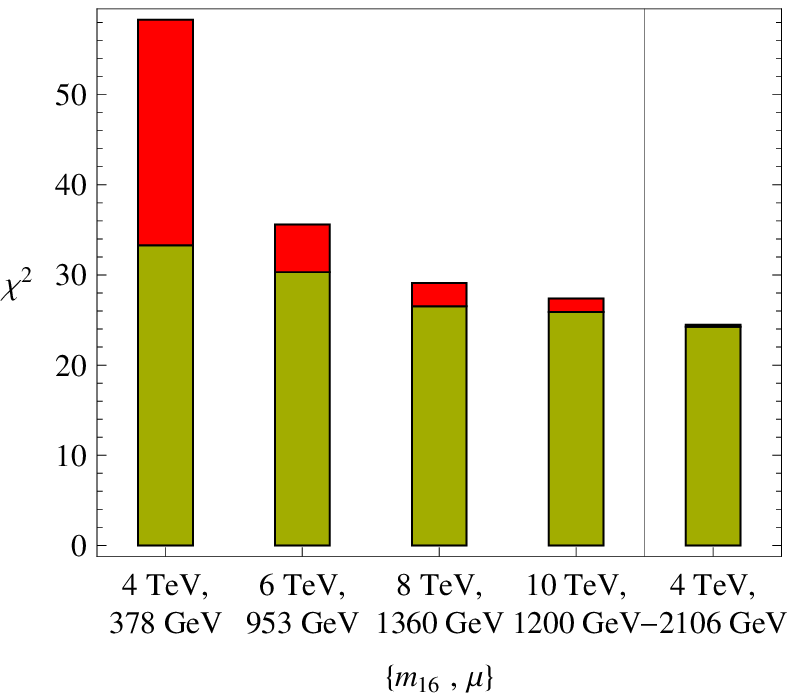}
\caption{Total $\chi^2$ (whole bar) and contribution of $B \to X_s\gamma$ to the $\chi^2$ (red bar) 
for the best fits with positive $\mu$ and $m_{16}=4$, 6, 8 and 10 TeV, respectively, as well as the 
best fit with negative $\mu$.}
\label{fig:barchart}
}
\noindent Considering the discussion in the last section, it is apparent that for positive $\mu$, the tension 
between the three decays $B\to X_s\gamma$, $B\to X_s\ell^+\ell^-$ and $B_s\to \mu^+\mu^-$ can be 
relieved by raising the universal sfermion mass $m_{16}$ beyond 8\,TeV. 
This is demonstrated in Fig. \ref{fig:barchart}, showing the total $\chi^2$ and the contribution of 
$B\to X_s\gamma$ to the $\chi^2$ for increasing values of $m_{16}$. For negative $\mu$, fits with 
comparable $\chi^2$ can be achieved with lower values of $m_{16}$, but not with lighter squarks, as 
the lightest squark is still very heavy in these cases.

However, it is well-known that the supersymmetric solution to the gauge hierarchy problem requires 
light third generation sfermions. Therefore, light stops are favourable to reduce fine-tuning.
To show the amount of splitting between fermion and sfermion masses needed in the DR model,
we plot the lightest stop mass $m_{\tilde t_1}$ versus the total $\chi^2$ for all fits with positive as 
well as negative $\mu$ we obtained (see Fig. \ref{fig:scatter}). 
There is obviously a strong correlation between the stop mass and the quality of the fit, 
demonstrating that $m_{\tilde t_1}$ has to be at least as large as 1.8\,TeV to cure the problems with 
the three aforementioned decays. This is significantly heavier than the masses considered in \cite{DR05, DR06} 
and may be difficult to reconcile with naturalness.
\FIGURE{
\includegraphics[width=7cm]{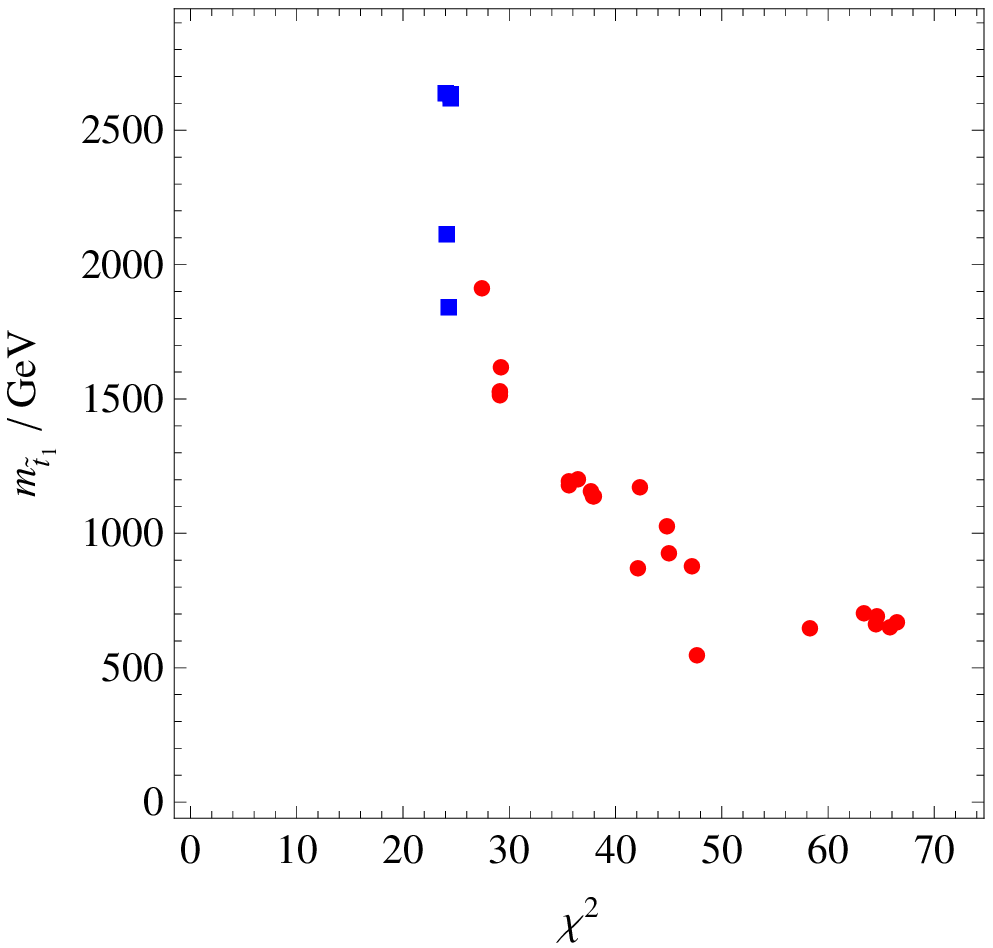}
\vspace{-1.5cm}
\caption{Total $\chi^2$ vs. the lightest stop mass for all obtained fits. Red circular points are fits 
with positive $\mu$, blue squares with negative $\mu$.}
\label{fig:scatter}
}

In addition, a number of problems of the model persist even for very large $m_{16}$. These are the 
issues related to the Yukawa textures: the up-quark mass, $V_{ub}$ and $\sin 2\beta$, and as a result 
also $\Delta M_s/\Delta M_d$ and $B^+\to\tau^+\nu$. This is why there are no points with 
$\chi^2 \lesssim 25$ in Fig. \ref{fig:scatter}. The reason for the much higher $\chi^2$ contribution 
from flavour conserving quantities as compared to \cite{DR05,DR06} is mainly due to updated experimental 
values and reduced experimental errors, especially in $|V_{us}|$, $|V_{ub}|$ and $m_b$.

For a given value of $m_{16}$, we found that successful fits could only be obtained in a limited range 
of allowed values for $\mu$. The preferred value of $\mu$ increased with increasing $m_{16}$, as can 
be seen for the best fit points in Fig. \ref{fig:barchart}. This fact could give rise to an additional 
problem.
Although we did not include the dark matter density as a constraint in our $\chi^2$ analysis, we 
conjecture that such large values of $\mu$ would give rise to a relic abundance of neutralinos 
incompatible with the WMAP measurements, for the following reason. Because of the extremely heavy 
sfermions, the dominant annihilation channel for the neutralino in this scenario is through an $s$-channel 
pseudoscalar Higgs. While this rate is already suppressed by the large $M_A$, the coupling of neutralinos 
to the pseudoscalar Higgs is additionally suppressed by large $\mu$. This would result in an
overabundance of relic neutralinos incompatible with observations.
Solving this problem by resonant neutralino annihilation with $m_{\tilde\chi^0}\approx M_A/2$, as was advocated
in an extensive analysis of dark matter in this class of models \cite{DRRR1,DRRR2},
is not possible because of the large $M_A$ and the small $M_{1/2}$ preferred by the fit.

\section{Conclusions} \label{sec:conclusions}

In this paper we have performed a detailed analysis of the SO(10) SUSY GUT model with $D_3$
family symmetry of \dmsk\ and Raby \cite{DR05,DR06}. 

This model is entirely specified in terms of 24 parameters. Once they are fixed, the whole
MSSM parameter space (including its SM subset) is predicted at low energies with the help 
of RGEs. The common dependence on the model parameters strongly correlates all the low energy
observables, in contrast with the direct consideration of the MSSM at the EW scale, where 
the CKM parameters and the fermion masses are fully independent of the SUSY particle
spectrum. We find that the DR model gives a satisfactory description of the quark and
lepton matrices as well as of the PMNS and CKM mixing matrices with one exception: the CKM
element $|V_{ub}|$ turns out to be significantly smaller than $|V_{ub}|_{\rm incl}$
extracted from inclusive tree-level decays and even smaller than $|V_{ub}|_{\rm excl}$.
The above findings are mostly a confirmation of previous studies of the model.

The main novelty of our study, with respect to similar analyses of SUSY GUT models found 
in the literature, is that we analyze simultaneously the mass spectra of quark and leptons, 
the CKM and PMNS mixing matrices, the SUSY mass spectrum and its implied corrections to 
the FC processes $B_{s} \to \mu^+ \mu^-$, $B \to X_s \gamma$, $B \to X_s \ell^+ \ell^-$, 
$B^+ \to \tau^+ \nu$ and the $B_{d,s} - \ov B_{d,s}$ mass differences $\D M_{d,s}$. The 
performance of the model is assessed by means of a global fit to the above mentioned 
observables.

The inclusion of the FC processes listed above turns out to be a crucial test of the mass
hierarchies predicted by the model for the SUSY spectrum. In fact, such hierarchies
unavoidably manifest themselves in loop corrections, and FC observables remain the most
sensitive probes of such corrections. Our analysis demonstrates that the simultaneous
description of all the FC processes listed above is a serious challenge for the DR model.
In view of the specific way this failure is realized, we suspect that this is a problem of a 
wider class of SUSY GUTs in which the presence of Yukawa unification implies $\tan \beta \simeq 50$,
unless non-minimal sources of flavour violation are introduced.

Our main message is the following one. To really assess the viability of models for flavour
parameters, it is essential not only to verify their ability to reproduce quark and lepton mass spectra 
and mixing matrices -- in itself an already notable achievement -- but also to test the
consistency with the data on available FC processes, since the latter have a simultaneous sensitivity 
to mixing matrices and new particles' spectra. In the DR model example, FC processes
are in fact the best probes available to the SUSY part of the spectrum, where information from 
direct detection is missing. It turns out that the DR model -- otherwise successful for quark and
lepton masses as well as for the CKM and PMNS matrices -- is challenged only when specifically tested 
in the simultaneous description of quark FCNC processes. The failure in the description of these data makes 
the viability of the DR model questionable from the present perspective, but hopefully offers insights on 
further lines of development along similar classes of models.

\section*{Acknowledgements}

We warmly acknowledge Radovan \dmsk\ and Stuart Raby for useful discussions and for a critical 
reading of the manuscript. This work has been supported in part by the 
Cluster of Excellence ``Origin and Structure of the Universe'' and by the German Bundesministerium 
f{\"u}r Bildung und Forschung under contract 05HT6WOA. D.G. also warmly acknowledges the
support of the A. von Humboldt Stiftung.

\section*{Note added}

During the completion of the present work, a new bound on the branching ratio for 
$B_s \to \mu^+ \mu^-$ has been presented at the HEP 2007 conference \cite{bsmumu-HEP07}. 
The latter results from a combined analysis of the CDF and D\O\ data and reads
\beqn
{\rm BR}(B_s \to \mu^+ \mu^-)_{{\rm CDF + D\text{\O}}} < 5.8 \times 10^{-8}, ~~~ \mbox{(95\% C.L.)}~.
\label{Bsmumu-new}
\eeqn
This bound represents a considerable improvement over the one given in eq. (\ref{bsmumu-exp}).

In addition, at the SUSY 2007 conference it was presented a new (preliminary) result on the 
BR$(B^+ \to \tau^+ \nu)$ from the BaBar collaboration \cite{btaunu-SUSY07}, which reads
\beqn
{\rm BR}(B^+ \to \tau^+ \nu)_{{\rm BaBar, prelim.}} = 
(1.2 \pm 0.4_{\rm stat} \pm 0.3_{\rm bkg} \pm 0.2_{\rm eff}) \times 10^{-4}~.
\label{btaunu-BaBar}
\eeqn
Performing the weighted average between the result in eq. (\ref{btaunu-BaBar}) and the
Belle result \cite{Belle-btaunu} one obtains
\beqn
{\rm BR}(B^+ \to \tau^+ \nu)_{{\rm new}} = (1.41 \pm 0.43) \times 10^{-4}~.
\label{Btaunu-new}
\eeqn
The new averages in eqs. (\ref{Bsmumu-new}) and (\ref{Btaunu-new}) further strengthen our conclusions.

\bibliography{biblio} 
\bibliographystyle{JHEP}

\TABLE[tp]{
\parbox{\textwidth}{
\centering
\input{table_parameters}
}
\caption{Input parameters for the fits presented in section \ref{sec:scenarios} (cf. Table \ref{tab:parameters}). 
Dimensionful quantities are given in units of GeV.}
\label{tab:fit-par}
}

\TABLE[tp]{
\input{tabm164000_mu+}
\caption{Fit results for the case $m_{16} = 4$ TeV, $\mu = 378$ GeV. The pull for the
$i^{\rm th}$ observable represents the square root of the corresponding entry in the
$\chi^2$ function (\ref{chi2}). Corresponding predictions are reported in Table
\ref{tab:predictions}. Dimensionful quantities are given in units of GeV.}
\label{tab:m164000_mu+}
}

\TABLE[tp]{
\input{tabm166000_mu+}
\caption{Fit results for the case $m_{16} = 6$ TeV, $\mu = 953$ GeV. Corresponding predictions are reported in 
Table \ref{tab:predictions}. Dimensionful quantities are given in units of GeV.}
\label{tab:m166000_mu+}
}

\TABLE[tp]{
\input{tabm1610000_mu+}
\caption{Fit results for the case $m_{16} = 10$ TeV, $\mu = 1200$ GeV. Corresponding predictions are 
reported in Table \ref{tab:predictions}. Dimensionful quantities are given in units of GeV.}
\label{tab:m1610000_mu+}
}

\TABLE[tp]{
\input{tabm164000_mu-}
\caption{Fit results for the case $m_{16} = 4$ TeV, $\mu = - 2106$ GeV. Corresponding predictions are 
reported in Table \ref{tab:predictions}. Dimensionful quantities are given in units of GeV.}
\label{tab:m164000_mu-}
}

\TABLE[ht]{
\parbox{\textwidth}{
\centering
\input{table_predictions}
}
\caption{Predictions for the scenarios presented in section \ref{sec:scenarios}. Masses are given in 
units of GeV. $\hat s_0$ is the zero position of the forward-backward asymmetry in $B\to X_s\ell^+\ell^-$ 
(cf. Section \protect\ref{sec:Bsll}). The quantity $\delta a_\mu^\text{SUSY}$ is the SUSY contribution to 
$a_\mu \equiv (g-2)_\mu/2$, which is currently measured to be $+27.6\times10^{-10}$ larger than the SM 
prediction \cite{g-2muon2,g-2muon3}, with an uncertainty of about $8\times10^{-10}$.}
\label{tab:predictions}
}

\end{document}

%% file: table_parameters.tex
\begin{tabular}{|l|l|l|l|l|}
\hline
$m_{16}  $  &  $4000$  &  $6000$  &  $10000$  &  $4000$  \\
$\mu  $  &  $378$  &  $953$  &  $1200$  &  $-2106$  \\
\hline\hline
$M_{1/2}  $  &  $147.3$  &  $145.6$  &  $146.7$  &  $229.9$  \\
$A_0  $  &  $-7787.4$  &  $-11924.0$  &  $-20070.0$  &  $-630.13$  \\
$\tan\beta  $  &  $49.9$  &  $48.8$  &  $48.7$  &  $49.3$  \\
$1/\alpha_G  $  &  $24.7$  &  $24.7$  &  $24.7$  &  $24.6$  \\
$M_G / 10^{16}  $  &  $2.45$  &  $3.11$  &  $4.8$  &  $4.96$  \\
$\epsilon_3 / \%  $  &  $-3.81$  &  $-4.12$  &  $-5.37$  &  $-5.6$  \\
$(m_{H_u}/m_{16})^2  $  &  $1.59$  &  $1.55$  &  $1.57$  &  $0.52$  \\
$(m_{H_d}/m_{16})^2  $  &  $1.86$  &  $1.79$  &  $1.8$  &  $1.0$  \\
$M_{R_1} / 10^{10}  $  &  $1.059$  &  $1.05$  &  $1.072$  &  $0.8999$  \\
$M_{R_2} / 10^{10}  $  &  $-74.85$  &  $-70.51$  &  $-71.93$  &  $-66.94$  \\
$M_{R_3} / 10^{10}  $  &  $3244.0$  &  $3053.0$  &  $3069.0$  &  $2718.0$  \\
$\lambda  $  &  $0.618$  &  $0.583$  &  $0.582$  &  $0.578$  \\
$\epsilon  $  &  $0.0473$  &  $0.048$  &  $0.0477$  &  $0.048$  \\
$\epsilon'  $  &  $-0.0034$  &  $-0.00338$  &  $-0.00342$  &  $-0.00356$  \\
$|\rho|  $  &  $0.0566$  &  $0.0584$  &  $0.0574$  &  $0.0567$  \\
$|\sigma|  $  &  $1.02$  &  $1.03$  &  $1.03$  &  $0.987$  \\
$|\tilde\epsilon|  $  &  $0.00964$  &  $0.00957$  &  $0.00967$  &  $0.00999$  \\
$|\xi|  $  &  $0.134$  &  $0.134$  &  $0.134$  &  $0.14$  \\
$\arg\rho  $  &  $3.93$  &  $3.92$  &  $3.94$  &  $3.94$  \\
$\arg\sigma  $  &  $0.641$  &  $0.617$  &  $0.622$  &  $0.751$  \\
$\arg\tilde\epsilon  $  &  $0.484$  &  $0.492$  &  $0.491$  &  $0.505$  \\
$\arg\xi  $  &  $3.61$  &  $3.6$  &  $3.61$  &  $3.58$  \\
\hline
\end{tabular}

%% file: tabm164000_mu+.tex
\begin{tabular}{|l|l|l|l|}
\hline
Observable  &  Exp. value  &  Fit value  &  Pull ($\sigma$)  \\
\hline\hline
$M_W$  &  80.403  &  80.4  &  0.1  \\
$M_Z$  &  91.1876  &  90.6  &  1.3  \\
$G_\text{F} \times 10^{5}$  &  1.16637  &  1.16  &  0.7  \\
$1/\alpha_\text{em}$  &  137.036  &  136.4  &  0.9  \\
$\alpha_s(M_Z)$  &  0.1176  &  0.115  &  1.1  \\
$M_t$  &  170.9  &  171.4  &  0.2  \\
$m_b(m_b)$  &  4.2  &  4.31  &  1.5  \\
$m_c(m_b)$  &  1.25  &  1.15  &  1.2  \\
$m_s(2\,\text{GeV})$  &  0.095  &  0.107  &  0.5  \\
$m_d(2\,\text{GeV})$  &  0.005  &  0.00741  &  1.2  \\
$m_u(2\,\text{GeV})$  &  0.00225  &  0.00462  &  \textbf{3.2}  \\
$M_\tau$  &  1.777  &  1.77  &  0.4  \\
$M_\mu$  &  0.10566  &  0.106  &  0.1  \\
$M_e$  &  0.000511  &  0.000511  &  0.0  \\
$|V_{us}|$  &  0.2258  &  0.225  &  0.6  \\
$|V_{ub}| \times 10^{3}$  &  4.1  &  3.26  &  \textbf{2.1}  \\
$|V_{cb}|$  &  0.0416  &  0.0417  &  0.1  \\
$\sin 2\beta$  &  0.675  &  0.637  &  1.4  \\
$\Delta m_{31}^2 \times 10^{21}$  &  2.6  &  2.6  &  0.0  \\
$\Delta m_{21}^2 \times 10^{23}$  &  7.9  &  7.9  &  0.0  \\
$\sin^2 2\theta_{12}$  &  0.852  &  0.85  &  0.1  \\
$\sin^2 2\theta_{23}$  &  0.996  &  1.0  &  0.2  \\
$\epsilon_K \times 10^{3}$  &  2.229  &  2.32  &  0.4  \\
\hline
$\text{BR} (B \to X_s \gamma) \times 10^{4}$  &  3.55  &  0.885  &  \textbf{5.0}  \\
$\text{BR} (B \to X_s \ell^+\ell^-) \times 10^{6}$  &  1.6  &  1.8  &  0.3  \\
$\Delta M_s / \Delta M_d$  &  35.05  &  29.8  &  1.4  \\
$\text{BR} (B^+ \to \tau^+\nu) \times 10^{4}$  &  1.31  &  0.336  &  \textbf{2.0}  \\
\hline
\multicolumn{3}{|r}{total $\chi^2$:}  &  \textbf{58.3} \\
\hline
\end{tabular}

%% file: tabm166000_mu+.tex
\begin{tabular}{|l|l|l|l|}
\hline
Observable  &  Exp. value  &  Fit value  &  Pull ($\sigma$)  \\
\hline\hline
$M_W$  &  80.403  &  80.5  &  0.2  \\
$M_Z$  &  91.1876  &  90.6  &  1.2  \\
$G_\text{F} \times 10^{5}$  &  1.16637  &  1.16  &  0.5  \\
$1/\alpha_\text{em}$  &  137.036  &  136.5  &  0.8  \\
$\alpha_s(M_Z)$  &  0.1176  &  0.116  &  0.5  \\
$M_t$  &  170.9  &  169.8  &  0.6  \\
$m_b(m_b)$  &  4.2  &  4.29  &  1.3  \\
$m_c(m_b)$  &  1.25  &  1.14  &  1.2  \\
$m_s(2\,\text{GeV})$  &  0.095  &  0.106  &  0.4  \\
$m_d(2\,\text{GeV})$  &  0.005  &  0.00727  &  1.1  \\
$m_u(2\,\text{GeV})$  &  0.00225  &  0.00465  &  \textbf{3.2}  \\
$M_\tau$  &  1.777  &  1.77  &  0.3  \\
$M_\mu$  &  0.10566  &  0.106  &  0.1  \\
$M_e$  &  0.000511  &  0.000511  &  0.0  \\
$|V_{us}|$  &  0.2258  &  0.225  &  0.6  \\
$|V_{ub}| \times 10^{3}$  &  4.1  &  3.26  &  \textbf{2.1}  \\
$|V_{cb}|$  &  0.0416  &  0.0417  &  0.1  \\
$\sin 2\beta$  &  0.675  &  0.638  &  1.4  \\
$\Delta m_{31}^2 \times 10^{21}$  &  2.6  &  2.6  &  0.0  \\
$\Delta m_{21}^2 \times 10^{23}$  &  7.9  &  7.9  &  0.0  \\
$\sin^2 2\theta_{12}$  &  0.852  &  0.852  &  0.0  \\
$\sin^2 2\theta_{23}$  &  0.996  &  0.997  &  0.1  \\
$\epsilon_K \times 10^{3}$  &  2.229  &  2.31  &  0.3  \\
\hline
$\text{BR} (B \to X_s \gamma) \times 10^{4}$  &  3.55  &  2.34  &  \textbf{2.3}  \\
$\text{BR} (B \to X_s \ell^+\ell^-) \times 10^{6}$  &  1.6  &  1.62  &  0.0  \\
$\Delta M_s / \Delta M_d$  &  35.05  &  30.0  &  1.4  \\
$\text{BR} (B^+ \to \tau^+\nu) \times 10^{4}$  &  1.31  &  0.398  &  1.9  \\
\hline
\multicolumn{3}{|r}{total $\chi^2$:}  &  \textbf{35.6} \\
\hline
\end{tabular}

%% file: tabm1610000_mu+.tex
\begin{tabular}{|l|l|l|l|}
\hline
Observable  &  Exp. value  &  Fit value  &  Pull ($\sigma$)  \\
\hline\hline
$M_W$  &  80.403  &  80.6  &  0.5  \\
$M_Z$  &  91.1876  &  90.7  &  1.1  \\
$G_\text{F} \times 10^{5}$  &  1.16637  &  1.16  &  0.3  \\
$1/\alpha_\text{em}$  &  137.036  &  136.8  &  0.4  \\
$\alpha_s(M_Z)$  &  0.1176  &  0.117  &  0.2  \\
$M_t$  &  170.9  &  170.6  &  0.2  \\
$m_b(m_b)$  &  4.2  &  4.22  &  0.3  \\
$m_c(m_b)$  &  1.25  &  1.14  &  1.2  \\
$m_s(2\,\text{GeV})$  &  0.095  &  0.107  &  0.5  \\
$m_d(2\,\text{GeV})$  &  0.005  &  0.00741  &  1.2  \\
$m_u(2\,\text{GeV})$  &  0.00225  &  0.00461  &  \textbf{3.1}  \\
$M_\tau$  &  1.777  &  1.78  &  0.1  \\
$M_\mu$  &  0.10566  &  0.106  &  0.1  \\
$M_e$  &  0.000511  &  0.000511  &  0.0  \\
$|V_{us}|$  &  0.2258  &  0.225  &  0.6  \\
$|V_{ub}| \times 10^{3}$  &  4.1  &  3.26  &  \textbf{2.1}  \\
$|V_{cb}|$  &  0.0416  &  0.0416  &  0.1  \\
$\sin 2\beta$  &  0.675  &  0.639  &  1.4  \\
$\Delta m_{31}^2 \times 10^{21}$  &  2.6  &  2.6  &  0.0  \\
$\Delta m_{21}^2 \times 10^{23}$  &  7.9  &  7.9  &  0.0  \\
$\sin^2 2\theta_{12}$  &  0.852  &  0.852  &  0.0  \\
$\sin^2 2\theta_{23}$  &  0.996  &  1.0  &  0.2  \\
$\epsilon_K \times 10^{3}$  &  2.229  &  2.33  &  0.4  \\
\hline
$\text{BR} (B \to X_s \gamma) \times 10^{4}$  &  3.55  &  2.86  &  1.3  \\
$\text{BR} (B \to X_s \ell^+\ell^-) \times 10^{6}$  &  1.6  &  1.62  &  0.0  \\
$\Delta M_s / \Delta M_d$  &  35.05  &  31.1  &  1.1  \\
$\text{BR} (B^+ \to \tau^+\nu) \times 10^{4}$  &  1.31  &  0.517  &  1.7  \\
\hline
\multicolumn{3}{|r}{total $\chi^2$:}  &  \textbf{27.4} \\
\hline
\end{tabular}

%% file: tabm164000_mu-.tex
\begin{tabular}{|l|l|l|l|}
\hline
Observable  &  Exp. value  &  Fit value  &  Pull ($\sigma$)  \\
\hline\hline
$M_W$  &  80.403  &  80.7  &  0.7  \\
$M_Z$  &  91.1876  &  90.7  &  1.1  \\
$G_\text{F} \times 10^{5}$  &  1.16637  &  1.17  &  0.2  \\
$1/\alpha_\text{em}$  &  137.036  &  136.9  &  0.3  \\
$\alpha_s(M_Z)$  &  0.1176  &  0.118  &  0.0  \\
$M_t$  &  170.9  &  170.5  &  0.2  \\
$m_b(m_b)$  &  4.2  &  4.19  &  0.1  \\
$m_c(m_b)$  &  1.25  &  1.14  &  1.2  \\
$m_s(2\,\text{GeV})$  &  0.095  &  0.0999  &  0.2  \\
$m_d(2\,\text{GeV})$  &  0.005  &  0.00716  &  1.1  \\
$m_u(2\,\text{GeV})$  &  0.00225  &  0.00446  &  \textbf{3.0}  \\
$M_\tau$  &  1.777  &  1.78  &  0.1  \\
$M_\mu$  &  0.10566  &  0.106  &  0.2  \\
$M_e$  &  0.000511  &  0.000511  &  0.1  \\
$|V_{us}|$  &  0.2258  &  0.224  &  1.2  \\
$|V_{ub}| \times 10^{3}$  &  4.1  &  3.26  &  \textbf{2.1}  \\
$|V_{cb}|$  &  0.0416  &  0.0416  &  0.0  \\
$\sin 2\beta$  &  0.675  &  0.64  &  1.3  \\
$\Delta m_{31}^2 \times 10^{21}$  &  2.6  &  2.6  &  0.0  \\
$\Delta m_{21}^2 \times 10^{23}$  &  7.9  &  7.9  &  0.0  \\
$\sin^2 2\theta_{12}$  &  0.852  &  0.851  &  0.0  \\
$\sin^2 2\theta_{23}$  &  0.996  &  0.996  &  0.0  \\
$\epsilon_K \times 10^{3}$  &  2.229  &  2.35  &  0.5  \\
\hline
$\text{BR} (B \to X_s \gamma) \times 10^{4}$  &  3.55  &  3.34  &  0.4  \\
$\text{BR} (B \to X_s \ell^+\ell^-) \times 10^{6}$  &  1.6  &  1.63  &  0.0  \\
$\Delta M_s / \Delta M_d$  &  35.05  &  31.4  &  1.0  \\
$\text{BR} (B^+ \to \tau^+\nu) \times 10^{4}$  &  1.31  &  0.59  &  1.5  \\
\hline
\multicolumn{3}{|r}{total $\chi^2$:}  &  \textbf{24.5} \\
\hline
\end{tabular}

%% file: table_predictions.tex
\begin{tabular}{|l|l|l|l|l|}
\hline
$m_{16}  $  &  $4000$  &  $6000$  &  $10000$  &  $4000$  \\
$\mu  $  &  $378$  &  $953$  &  $1200$  &  $-2106$  \\
\hline\hline
$\text{BR} (B_s \to \mu^+\mu^-) \times 10^{8}  $  &  $8.6$  &  $7.7$  &  $2.1$  &  $0.33$  \\
$\hat s_0$  &  0.022  &  0.13  &  0.14  &  0.16  \\
$\text{BR} (\mu \to e \gamma) \times 10^{13}  $  &  $0.36$  &  $0.021$  &  $0.0026$  &  $0.011$  \\
$\delta a_\mu^\text{SUSY} \times 10^{10}  $  &  $+5.8$  &  $+1.6$  &  $+0.52$  &  $-2.9$  \\
\hline
$M_{h_0}$    &  126  &  129  &  129  &  119  \\
$M_A$    &  507  &  559  &  842  &  1800  \\
$m_{\tilde t_1}$    &  640  &  1172  &  1903  &  2627  \\
$m_{\tilde b_1}$    &  895  &  1475  &  2366  &  2488  \\
$m_{\tilde \tau_1}$    &  1510  &  2419  &  3933  &  2931  \\
$m_{\tilde\chi^0_1}$    &  60  &  60  &  60  &  94  \\
$m_{\tilde\chi^+_1}$    &  115  &  119  &  120  &  189  \\
$m_{\tilde g}$    &  462 & 478 & 506 & 703  \\
\hline
\end{tabular}

%% file: AABuGuS.bbl
\providecommand{\href}[2]{#2}\begingroup\raggedright\begin{thebibliography}{10%
0}

\bibitem{DR05}
R.~Derm{\'i}{\v{s}}ek and S.~Raby, {\it {Bi-large neutrino mixing and CP
  violation in an SO(10) SUSY GUT for fermion masses}},  {\em Phys. Lett.} {\bf
  B622} (2005) 327--338, [\href{http://xxx.lanl.gov/abs/hep-ph/0507045}{{\tt
  hep-ph/0507045}}].

\bibitem{DR06}
R.~Derm{\'i}{\v{s}}ek, M.~Harada, and S.~Raby, {\it {SO(10) SUSY GUT for
  fermion masses: Lepton flavor and CP violation}},  {\em Phys. Rev.} {\bf D74}
  (2006) 035011, [\href{http://xxx.lanl.gov/abs/hep-ph/0606055}{{\tt
  hep-ph/0606055}}].

\bibitem{DR-D3symmetry}
R.~Derm{\'i}{\v{s}}ek and S.~Raby, {\it {Fermion masses and neutrino
  oscillations in SO(10) SUSY GUT with D(3) $\times$ U(1) family symmetry}},
  {\em Phys. Rev.} {\bf D62} (2000) 015007,
  [\href{http://xxx.lanl.gov/abs/hep-ph/9911275}{{\tt hep-ph/9911275}}].

\bibitem{FroggattNielsen}
C.~D. Froggatt and H.~B. Nielsen, {\it {Hierarchy of Quark Masses, Cabibbo
  Angles and CP Violation}},  {\em Nucl. Phys.} {\bf B147} (1979) 277.

\bibitem{Minkowski}
P.~Minkowski, {\it {$\mu \to e \gamma$ at a Rate of One Out of 1-Billion Muon
  Decays?}},  {\em Phys. Lett.} {\bf B67} (1977) 421.

\bibitem{see-saw1}
M. Gell-Mann, P. Ramond and R. Slansky, {\em Supergravity} (P. van
  Nieuwenhuizen and D.Z. Freedman eds.), North-Holland, Amsterdam, 1979, p.
  315; T. Yanagida, in {\em Proceedings of the Workshop on the unified theory
  and the baryon number of the universe} (O. Sawada and A. Sugamoto eds.), KEK
  report No. 79-18, Tsukuba, Japan, 1979, p. 95; S. L. Glashow, {\em The future
  of elementary particle physics}, in {\em Proceedings of the 1979 Carg{\`e}se
  Summer Institute on Quarks and Leptons} (M. L{\'e}vy et al. eds.), Plenum
  Press, New York, 1980, pp. 687-713.

\bibitem{see-saw2}
R.~N. Mohapatra and G.~Senjanovic, {\it {Neutrino mass and spontaneous parity
  nonconservation}},  {\em Phys. Rev. Lett.} {\bf 44} (1980) 912.

\bibitem{see-saw3}
P.~Ramond, {\it {The family group in grand unified theories}},
  \href{http://xxx.lanl.gov/abs/hep-ph/9809459}{{\tt hep-ph/9809459}}.

\bibitem{ATR}
R.~Ruiz~de Austri, R.~Trotta, and L.~Roszkowski, {\it {A Markov chain Monte
  Carlo analysis of the CMSSM}},  {\em JHEP} {\bf 05} (2006) 002,
  [\href{http://xxx.lanl.gov/abs/hep-ph/0602028}{{\tt hep-ph/0602028}}].

\bibitem{RAT}
L.~Roszkowski, R.~Ruiz~de Austri, and R.~Trotta, {\it {On the detectability of
  the CMSSM light Higgs boson at the Tevatron}},  {\em JHEP} {\bf 04} (2007)
  084, [\href{http://xxx.lanl.gov/abs/hep-ph/0611173}{{\tt hep-ph/0611173}}].

\bibitem{RAT2}
L.~Roszkowski, R.~Ruiz~de Austri, and R.~Trotta, {\it {Implications for the
  constrained MSSM from a new prediction for $b \to s \gamma$}},
  \href{http://xxx.lanl.gov/abs/0705.2012}{{\tt 0705.2012}}.

\bibitem{Ellis2007}
J.~R. Ellis, S.~Heinemeyer, K.~A. Olive, A.~M. Weber, and G.~Weiglein, {\it
  {The Supersymmetric Parameter Space in Light of B-physics Observables and
  Electroweak Precision Data}},  \href{http://xxx.lanl.gov/abs/0706.0652}{{\tt
  0706.0652}}.

\bibitem{Hisano95}
J.~Hisano, T.~Moroi, K.~Tobe, and M.~Yamaguchi, {\it {Lepton-Flavor Violation
  via Right-Handed Neutrino Yukawa Couplings in Supersymmetric Standard
  Model}},  {\em Phys. Rev.} {\bf D53} (1996) 2442--2459,
  [\href{http://xxx.lanl.gov/abs/hep-ph/9510309}{{\tt hep-ph/9510309}}].

\bibitem{AKLR05}
S.~Antusch, J.~Kersten, M.~Lindner, M.~Ratz, and M.~A. Schmidt, {\it {Running
  neutrino mass parameters in see-saw scenarios}},  {\em JHEP} {\bf 03} (2005)
  024, [\href{http://xxx.lanl.gov/abs/hep-ph/0501272}{{\tt hep-ph/0501272}}].

\bibitem{Petcov03}
S.~T. Petcov, S.~Profumo, Y.~Takanishi, and C.~E. Yaguna, {\it {Charged lepton
  flavor violating decays: Leading logarithmic approximation versus full RG
  results}},  {\em Nucl. Phys.} {\bf B676} (2004) 453--480,
  [\href{http://xxx.lanl.gov/abs/hep-ph/0306195}{{\tt hep-ph/0306195}}].

\bibitem{AKLR02}
S.~Antusch, J.~Kersten, M.~Lindner, and M.~Ratz, {\it {Neutrino mass matrix
  running for non-degenerate see-saw scales}},  {\em Phys. Lett.} {\bf B538}
  (2002) 87--95, [\href{http://xxx.lanl.gov/abs/hep-ph/0203233}{{\tt
  hep-ph/0203233}}].

\bibitem{MartinVaughn}
S.~P. Martin and M.~T. Vaughn, {\it {Two loop renormalization group equations
  for soft supersymmetry breaking couplings}},  {\em Phys. Rev.} {\bf D50}
  (1994) 2282, [\href{http://xxx.lanl.gov/abs/hep-ph/9311340}{{\tt
  hep-ph/9311340}}].

\bibitem{MFV}
G.~D'Ambrosio, G.~F. Giudice, G.~Isidori, and A.~Strumia, {\it {Minimal flavour
  violation: An effective field theory approach}},  {\em Nucl. Phys.} {\bf
  B645} (2002) 155--187, [\href{http://xxx.lanl.gov/abs/hep-ph/0207036}{{\tt
  hep-ph/0207036}}].

\bibitem{RosiekFR}
J.~Rosiek, {\it {Complete set of Feynman rules for the Minimal Supersymmetric
  extension of the Standard Model}},  {\em Phys. Rev.} {\bf D41} (1990) 3464.
  Erratum {\tt [hep-ph/9511250]}.

\bibitem{PBMZ}
D.~M. Pierce, J.~A. Bagger, K.~T. Matchev, and R.-J. Zhang, {\it {Precision
  corrections in the Minimal Supersymmetric Standard Model}},  {\em Nucl.
  Phys.} {\bf B491} (1997) 3--67,
  [\href{http://xxx.lanl.gov/abs/hep-ph/9606211}{{\tt hep-ph/9606211}}].

\bibitem{ChankowskiOSH}
P.~H. Chankowski, S.~Pokorski, and J.~Rosiek, {\it {Complete on-shell
  renormalization scheme for the minimal supersymmetric Higgs sector}},  {\em
  Nucl. Phys.} {\bf B423} (1994) 437--496,
  [\href{http://xxx.lanl.gov/abs/hep-ph/9303309}{{\tt hep-ph/9303309}}].

\bibitem{FeynHiggs1}
S.~Heinemeyer, W.~Hollik, and G.~Weiglein, {\it {FeynHiggs: A program for the
  calculation of the masses of the neutral CP-even Higgs bosons in the MSSM}},
  {\em Comput. Phys. Commun.} {\bf 124} (2000) 76--89,
  [\href{http://xxx.lanl.gov/abs/hep-ph/9812320}{{\tt hep-ph/9812320}}].

\bibitem{FeynHiggs2}
S.~Heinemeyer, W.~Hollik, and G.~Weiglein, {\it {The masses of the neutral
  CP-even Higgs bosons in the MSSM: Accurate analysis at the two-loop level}},
  {\em Eur. Phys. J.} {\bf C9} (1999) 343--366,
  [\href{http://xxx.lanl.gov/abs/hep-ph/9812472}{{\tt hep-ph/9812472}}].

\bibitem{FeynHiggs3}
G.~Degrassi, S.~Heinemeyer, W.~Hollik, P.~Slavich, and G.~Weiglein, {\it
  {Towards high-precision predictions for the MSSM Higgs sector}},  {\em Eur.
  Phys. J.} {\bf C28} (2003) 133--143,
  [\href{http://xxx.lanl.gov/abs/hep-ph/0212020}{{\tt hep-ph/0212020}}].

\bibitem{FeynHiggs4}
M.~Frank {\em et.~al.}, {\it {The Higgs boson masses and mixings of the complex
  MSSM in the Feynman-diagrammatic approach}},
  \href{http://xxx.lanl.gov/abs/hep-ph/0611326}{{\tt hep-ph/0611326}}.

\bibitem{CGNW}
M.~Carena, D.~Garcia, U.~Nierste, and C.~E.~M. Wagner, {\it {Effective
  Lagrangian for the $\bar{t} b H^+$ interaction in the MSSM and charged Higgs
  phenomenology}},  {\em Nucl. Phys.} {\bf B577} (2000) 88--120,
  [\href{http://xxx.lanl.gov/abs/hep-ph/9912516}{{\tt hep-ph/9912516}}].

\bibitem{BRP}
T.~Bla{\v{z}}ek, S.~Raby, and S.~Pokorski, {\it {Finite supersymmetric
  threshold corrections to CKM matrix elements in the large $\tan\beta$
  regime}},  {\em Phys. Rev.} {\bf D52} (1995) 4151--4158,
  [\href{http://xxx.lanl.gov/abs/hep-ph/9504364}{{\tt hep-ph/9504364}}].

\bibitem{BCRSbig}
A.~J. Buras, P.~H. Chankowski, J.~Rosiek, and L.~Slawianowska, {\it {$\Delta
  M_{d,s}$, $B_{d,s}^0 \to \mu^+ \mu^-$ and $B \to X_s \gamma$ in supersymmetry
  at large $\tan(\beta)$}},  {\em Nucl. Phys.} {\bf B659} (2003) 3,
  [\href{http://xxx.lanl.gov/abs/hep-ph/0210145}{{\tt hep-ph/0210145}}].

\bibitem{ChankowskiWasowicz}
P.~H. Chankowski and P.~Wasowicz, {\it {Low energy threshold corrections to
  neutrino masses and mixing angles}},  {\em Eur. Phys. J.} {\bf C23} (2002)
  249--258, [\href{http://xxx.lanl.gov/abs/hep-ph/0110237}{{\tt
  hep-ph/0110237}}].

\bibitem{BurasPLB}
A.~J. Buras, {\it {Relations between $\Delta M_{s,d}$ and $B_{s,d} \to \mu \ov
  \mu$ in models with minimal flavour violation}},  {\em Phys. Lett.} {\bf
  B566} (2003) 115--119, [\href{http://xxx.lanl.gov/abs/hep-ph/0303060}{{\tt
  hep-ph/0303060}}].

\bibitem{BBGT}
M.~Blanke, A.~J. Buras, D.~Guadagnoli, and C.~Tarantino, {\it {Minimal flavour
  violation waiting for precise measurements of $\Delta M_s$, $S_{\psi \phi}$,
  $A_{\rm SL}^s$, $|V_{ub}|$, $\gamma$ and $B_{s,d}^0 \to \mu^+ \mu^-$}},  {\em
  JHEP} {\bf 10} (2006) 003,
  [\href{http://xxx.lanl.gov/abs/hep-ph/0604057}{{\tt hep-ph/0604057}}].

\bibitem{CDF-bsmumu}
{\sf http://www-cdf.fnal.gov/physics/new/bottom/060316.blessed-bsmumu3} and CDF
  Public note 8176.

\bibitem{gaur}
S.~R. Choudhury and N.~Gaur, {\it {Dileptonic decay of $B_s$ meson in SUSY
  models with large $\tan(\beta)$}},  {\em Phys. Lett.} {\bf B451} (1999)
  86--92, [\href{http://xxx.lanl.gov/abs/hep-ph/9810307}{{\tt
  hep-ph/9810307}}].

\bibitem{BabuKolda}
K.~S. Babu and C.~F. Kolda, {\it {Higgs-mediated $B_0 \to \mu^+ \mu^-$ in
  minimal supersymmetry}},  {\em Phys. Rev. Lett.} {\bf 84} (2000) 228--231,
  [\href{http://xxx.lanl.gov/abs/hep-ph/9909476}{{\tt hep-ph/9909476}}].

\bibitem{IsidoriRetico}
G.~Isidori and A.~Retico, {\it {Scalar flavour-changing neutral currents in the
  large- $\tan(\beta)$ limit}},  {\em JHEP} {\bf 11} (2001) 001,
  [\href{http://xxx.lanl.gov/abs/hep-ph/0110121}{{\tt hep-ph/0110121}}].

\bibitem{CMW}
M.~Carena, A.~Menon, and C.~E.~M. Wagner, {\it {Challenges for MSSM Higgs
  searches at Hadron Colliders}},
  \href{http://xxx.lanl.gov/abs/0704.1143}{{\tt 0704.1143}}.

\bibitem{Bparams-lattice-DB}
D.~Becirevic, V.~Gimenez, G.~Martinelli, M.~Papinutto, and J.~Reyes, {\it
  {B-parameters of the complete set of matrix elements of $\Delta B = 2$
  operators from the lattice}},  {\em JHEP} {\bf 04} (2002) 025,
  [\href{http://xxx.lanl.gov/abs/hep-lat/0110091}{{\tt hep-lat/0110091}}].

\bibitem{FGH}
A.~Freitas, E.~Gasser, and U.~Haisch, {\it {Supersymmetric large $\tan(\beta)$
  corrections to $\Delta M_{d,s}$ and $B_{d,s} \to \mu^+ \mu^-$ revisited}},
  \href{http://xxx.lanl.gov/abs/hep-ph/0702267}{{\tt hep-ph/0702267}}. See also
  M.~Gorbahn, S.~J{\"{a}}ger, U.~Nierste and S.~Trine. In preparation.

\bibitem{DeltaMs-exp}
{\bf CDF} Collaboration, A.~Abulencia {\em et.~al.}, {\it Observation of
  {$B_s^0$}-{$\bar{B}_s^0$} oscillations},  {\em Phys. Rev. Lett.} {\bf 97}
  (2006) 242003, [\href{http://xxx.lanl.gov/abs/hep-ex/0609040}{{\tt
  hep-ex/0609040}}].

\bibitem{UTfit}
{{\bf UTfit} website: {\sf http://www.utfit.org}}.

\bibitem{CKMfitter}
{{\bf CKMfitter} website: {\sf http://ckmfitter.in2p3.fr}}.

\bibitem{BCRS-PL}
A.~J. Buras, P.~H. Chankowski, J.~Rosiek, and L.~Slawianowska, {\it
  {Correlation between $\Delta M_s$ and $B_{s,d}^0 \to \mu^+ \mu^-$ in
  supersymmetry at large $\tan(\beta)$}},  {\em Phys. Lett.} {\bf B546} (2002)
  96--107, [\href{http://xxx.lanl.gov/abs/hep-ph/0207241}{{\tt
  hep-ph/0207241}}].

\bibitem{Belle-bsgamma}
{\bf Belle} Collaboration, P.~Koppenburg {\em et.~al.}, {\it {An inclusive
  measurement of the photon energy spectrum in $b \to s \gamma$ decays}},  {\em
  Phys. Rev. Lett.} {\bf 93} (2004) 061803,
  [\href{http://xxx.lanl.gov/abs/hep-ex/0403004}{{\tt hep-ex/0403004}}].

\bibitem{Babar-bsgamma}
{\bf BaBar} Collaboration, B.~Aubert {\em et.~al.}, {\it {Measurement of the
  branching fraction and photon energy moments of $B \to X_s \gamma$ and
  $A_{\rm CP}(B \to X_{s+d} \gamma)$}},  {\em Phys. Rev. Lett.} {\bf 97} (2006)
  171803, [\href{http://xxx.lanl.gov/abs/hep-ex/0607071}{{\tt
  hep-ex/0607071}}].

\bibitem{HFAG-bsgamma}
{\bf Heavy Flavor Averaging Group (HFAG)} Collaboration, E.~Barberio {\em
  et.~al.}, {\it {Averages of b-hadron properties at the end of 2005}},
  \href{http://xxx.lanl.gov/abs/hep-ex/0603003}{{\tt hep-ex/0603003}}.

\bibitem{Misiak-NNLO}
M.~Misiak {\em et.~al.}, {\it {The first estimate of ${\rm B}(\ov B \to X_s
  \gamma)$ at O$(\alpha_s^2)$}},  {\em Phys. Rev. Lett.} {\bf 98} (2007)
  022002, [\href{http://xxx.lanl.gov/abs/hep-ph/0609232}{{\tt
  hep-ph/0609232}}].

\bibitem{becher-neubert}
T.~Becher and M.~Neubert, {\it {Analysis of $Br(B \to X_s \gamma)$ at NNLO with
  a cut on photon energy}},  {\em Phys. Rev. Lett.} {\bf 98} (2007) 022003,
  [\href{http://xxx.lanl.gov/abs/hep-ph/0610067}{{\tt hep-ph/0610067}}].

\bibitem{BMU}
C.~Bobeth, M.~Misiak, and J.~Urban, {\it {Matching conditions for $b \to s
  \gamma$ and $b \to s$ gluon in extensions of the standard model}},  {\em
  Nucl. Phys.} {\bf B567} (2000) 153--185,
  [\href{http://xxx.lanl.gov/abs/hep-ph/9904413}{{\tt hep-ph/9904413}}].

\bibitem{bsgammaMSSM1}
G.~Degrassi, P.~Gambino, and G.~F. Giudice, {\it {$B \to X_s \gamma$ in
  supersymmetry: Large contributions beyond the leading order}},  {\em JHEP}
  {\bf 12} (2000) 009, [\href{http://xxx.lanl.gov/abs/hep-ph/0009337}{{\tt
  hep-ph/0009337}}].

\bibitem{bsgammaMSSM2}
M.~Carena, D.~Garcia, U.~Nierste, and C.~E.~M. Wagner, {\it {$b \to s \gamma$
  and supersymmetry with large $\tan\beta$}},  {\em Phys. Lett.} {\bf B499}
  (2001) 141--146, [\href{http://xxx.lanl.gov/abs/hep-ph/0010003}{{\tt
  hep-ph/0010003}}].

\bibitem{BMU-NNLO-bsll}
C.~Bobeth, M.~Misiak, and J.~Urban, {\it {Photonic penguins at two loops and
  $m_t$-dependence of ${\rm BR}(B \to X_s \ell^+ \ell^-)$}},  {\em Nucl. Phys.}
  {\bf B574} (2000) 291--330,
  [\href{http://xxx.lanl.gov/abs/hep-ph/9910220}{{\tt hep-ph/9910220}}].

\bibitem{Asatryan-bsll1}
H.~H. Asatryan, H.~M. Asatrian, C.~Greub, and M.~Walker, {\it {Calculation of
  two loop virtual corrections to $b \to s \ell^+ \ell^-$ in the standard
  model}},  {\em Phys. Rev.} {\bf D65} (2002) 074004,
  [\href{http://xxx.lanl.gov/abs/hep-ph/0109140}{{\tt hep-ph/0109140}}].

\bibitem{Asatryan-bsll2}
H.~H. Asatryan, H.~M. Asatrian, C.~Greub, and M.~Walker, {\it {Complete gluon
  bremsstrahlung corrections to the process $b \to s \ell^+ \ell^-$}},  {\em
  Phys. Rev.} {\bf D66} (2002) 034009,
  [\href{http://xxx.lanl.gov/abs/hep-ph/0204341}{{\tt hep-ph/0204341}}].

\bibitem{GHIY}
A.~Ghinculov, T.~Hurth, G.~Isidori, and Y.~P. Yao, {\it {The rare decay $B \to
  X_s \ell^+ \ell^-$ to NNLL precision for arbitrary dilepton invariant mass}},
   {\em Nucl. Phys.} {\bf B685} (2004) 351--392,
  [\href{http://xxx.lanl.gov/abs/hep-ph/0312128}{{\tt hep-ph/0312128}}].

\bibitem{GGH}
P.~Gambino, M.~Gorbahn, and U.~Haisch, {\it {Anomalous dimension matrix for
  radiative and rare semileptonic B decays up to three loops}},  {\em Nucl.
  Phys.} {\bf B673} (2003) 238--262,
  [\href{http://xxx.lanl.gov/abs/hep-ph/0306079}{{\tt hep-ph/0306079}}].

\bibitem{BGGH}
C.~Bobeth, P.~Gambino, M.~Gorbahn, and U.~Haisch, {\it {Complete NNLO QCD
  analysis of $\ov B \to X_s \ell^+ \ell^-$ and higher order electroweak
  effects}},  {\em JHEP} {\bf 04} (2004) 071,
  [\href{http://xxx.lanl.gov/abs/hep-ph/0312090}{{\tt hep-ph/0312090}}].

\bibitem{HLMW-NNLO-bsll}
T.~Huber, E.~Lunghi, M.~Misiak, and D.~Wyler, {\it {Electromagnetic Logarithms
  in $B \to X_s \ell^+ \ell^-$}},  {\em Nucl. Phys.} {\bf B740} (2006)
  105--137, [\href{http://xxx.lanl.gov/abs/hep-ph/0512066}{{\tt
  hep-ph/0512066}}].

\bibitem{Beneke1}
M.~Beneke, T.~Feldmann, and D.~Seidel, {\it {Systematic approach to exclusive
  $B \to V \ell^+ \ell^-$, $V \gamma$ decays}},  {\em Nucl. Phys.} {\bf B612}
  (2001) 25--58, [\href{http://xxx.lanl.gov/abs/hep-ph/0106067}{{\tt
  hep-ph/0106067}}].

\bibitem{Beneke2}
M.~Beneke, T.~Feldmann, and D.~Seidel, {\it {Exclusive radiative and
  electroweak $b \to d$ and $b \to s$ penguin decays at NLO}},  {\em Eur. Phys.
  J.} {\bf C41} (2005) 173--188,
  [\href{http://xxx.lanl.gov/abs/hep-ph/0412400}{{\tt hep-ph/0412400}}].

\bibitem{Misiak-bsll-SM}
M.~Misiak, {\it {The $b \to s e^+ e^-$ and $b \to s \gamma$ decays with
  next-to-leading logarithmic QCD corrections}},  {\em Nucl. Phys.} {\bf B393}
  (1993) 23--45.

\bibitem{BurasMunz-bsll-SM}
A.~J. Buras and M.~M{\"{u}}nz, {\it {Effective Hamiltonian for $B \to X_s e^+
  e^-$ beyond leading logarithms in the NDR and HV schemes}},  {\em Phys. Rev.}
  {\bf D52} (1995) 186--195,
  [\href{http://xxx.lanl.gov/abs/hep-ph/9501281}{{\tt hep-ph/9501281}}].

\bibitem{hiller-krueger}
G.~Hiller and F.~Kr{\"{u}}ger, {\it {More model-independent analysis of $b \to
  s$ processes}},  {\em Phys. Rev.} {\bf D69} (2004) 074020,
  [\href{http://xxx.lanl.gov/abs/hep-ph/0310219}{{\tt hep-ph/0310219}}].

\bibitem{BBL}
G.~Buchalla, A.~J. Buras, and M.~E. Lautenbacher, {\it {Weak decays beyond
  leading logarithms}},  {\em Rev. Mod. Phys.} {\bf 68} (1996) 1125--1144,
  [\href{http://xxx.lanl.gov/abs/hep-ph/9512380}{{\tt hep-ph/9512380}}]. See
  also references therein.

\bibitem{BBE-NNLO}
C.~Bobeth, A.~J. Buras, and T.~Ewerth, {\it {$\ov B \to X_s \ell^+ \ell^-$ in
  the MSSM at NNLO}},  {\em Nucl. Phys.} {\bf B713} (2005) 522--554,
  [\href{http://xxx.lanl.gov/abs/hep-ph/0409293}{{\tt hep-ph/0409293}}].

\bibitem{ALGH}
A.~Ali, E.~Lunghi, C.~Greub, and G.~Hiller, {\it {Improved model-independent
  analysis of semileptonic and radiative rare B decays}},  {\em Phys. Rev.}
  {\bf D66} (2002) 034002, [\href{http://xxx.lanl.gov/abs/hep-ph/0112300}{{\tt
  hep-ph/0112300}}].

\bibitem{Burdman}
G.~Burdman, {\it {Short distance coefficients and the vanishing of the lepton
  asymmetry in $B \to V \ell^+ \ell^-$}},  {\em Phys. Rev.} {\bf D57} (1998)
  4254--4257, [\href{http://xxx.lanl.gov/abs/hep-ph/9710550}{{\tt
  hep-ph/9710550}}].

\bibitem{Asatrian-bsll3}
H.~H. Asatrian, H.~M. Asatrian, C.~Greub, and M.~Walker, {\it {Two-loop virtual
  corrections to $B \to X_s \ell^+ \ell^-$ in the Standard Model}},  {\em Phys.
  Lett.} {\bf B507} (2001) 162--172,
  [\href{http://xxx.lanl.gov/abs/hep-ph/0103087}{{\tt hep-ph/0103087}}].

\bibitem{Asatrian-bsll4}
H.~M. Asatrian, K.~Bieri, C.~Greub, and A.~Hovhannisyan, {\it {NNLL corrections
  to the angular distribution and to the forward-backward asymmetries in $B \to
  X_s \ell^+ \ell^-$}},  {\em Phys. Rev.} {\bf D66} (2002) 094013,
  [\href{http://xxx.lanl.gov/abs/hep-ph/0209006}{{\tt hep-ph/0209006}}].

\bibitem{Belle-bsll}
{\bf Belle} Collaboration, M.~Iwasaki {\em et.~al.}, {\it {Improved measurement
  of the electroweak penguin process $B \to X_s \ell^+ \ell^-$}},  {\em Phys.
  Rev.} {\bf D72} (2005) 092005,
  [\href{http://xxx.lanl.gov/abs/hep-ex/0503044}{{\tt hep-ex/0503044}}].

\bibitem{Babar-bsll}
{\bf BaBar} Collaboration, B.~Aubert {\em et.~al.}, {\it {Measurement of the $B
  \to X_s \ell^+ \ell^-$ branching fraction with a sum over exclusive modes}},
  {\em Phys. Rev. Lett.} {\bf 93} (2004) 081802,
  [\href{http://xxx.lanl.gov/abs/hep-ex/0404006}{{\tt hep-ex/0404006}}].

\bibitem{Belle-AFB}
A.~Ishikawa {\em et.~al.}, {\it {Measurement of forward-backward asymmetry and
  Wilson coefficients in $B \to K^* \ell^+ \ell^-$}},  {\em Phys. Rev. Lett.}
  {\bf 96} (2006) 251801, [\href{http://xxx.lanl.gov/abs/hep-ex/0603018}{{\tt
  hep-ex/0603018}}].

\bibitem{LPV}
E.~Lunghi, W.~Porod, and O.~Vives, {\it {Analysis of enhanced $\tan(\beta)$
  corrections in MFV GUT scenarios}},  {\em Phys. Rev.} {\bf D74} (2006)
  075003, [\href{http://xxx.lanl.gov/abs/hep-ph/0605177}{{\tt
  hep-ph/0605177}}].

\bibitem{BPSW}
A.~J. Buras, A.~Poschenrieder, M.~Spranger, and A.~Weiler, {\it {The impact of
  universal extra dimensions on $B \to X_s \gamma$, $B \to X_s {\rm gluon}$, $B
  \to X_s \mu^+ \mu^-$, $K_L \to \pi^0 e^+ e^-$, and $\epsilon'/\epsilon$}},
  {\em Nucl. Phys.} {\bf B678} (2004) 455--490,
  [\href{http://xxx.lanl.gov/abs/hep-ph/0306158}{{\tt hep-ph/0306158}}].

\bibitem{Ikado}
K. Ikado, talk presented at FPCP 2006 (9-12 April 2006, Vancouver, Canada) {\sf
  http://fpcp2006.triumf.ca}.

\bibitem{HFAG}
{Heavy Flavor Averaging Group:\\ {\sf http://www.slac.stanford.edu/xorg/hfag}}.

\bibitem{PDBook}
W.-M. {Yao} {\em et.~al.}, {\it {Review of Particle Physics}},  {\em {Journal
  of Physics G}} {\bf 33} (2006) 1+.

\bibitem{lattice}
S.~Hashimoto, {\it {Recent results from lattice calculations}},  {\em Int. J.
  Mod. Phys.} {\bf A20} (2005) 5133--5144,
  [\href{http://xxx.lanl.gov/abs/hep-ph/0411126}{{\tt hep-ph/0411126}}].

\bibitem{Belle-btaunu}
K.~Ikado {\em et.~al.}, {\it {Evidence of the purely leptonic decay $B^- \to
  \tau^- \ov\nu_\tau$}},  {\em Phys. Rev. Lett.} {\bf 97} (2006) 251802,
  [\href{http://xxx.lanl.gov/abs/hep-ex/0604018}{{\tt hep-ex/0604018}}].

\bibitem{Babar-btaunu}
{\bf BaBar} Collaboration, B.~Aubert {\em et.~al.}, {\it {A search for $B^+ \to
  \tau^+ \nu$ recoiling against $B^- \to D^0 \ell^- \ov\nu_\ell X$}},
  \href{http://xxx.lanl.gov/abs/hep-ex/0608019}{{\tt hep-ex/0608019}}.

\bibitem{UTfit-btaunu}
{\bf UTfit} Collaboration, M.~Bona {\em et.~al.}, {\it {The unitarity triangle
  fit in the standard model and hadronic parameters from lattice QCD: A
  reappraisal after the measurements of $\Delta m_s$ and BR$(B \to \tau
  \nu_\tau)$}},  {\em JHEP} {\bf 10} (2006) 081,
  [\href{http://xxx.lanl.gov/abs/hep-ph/0606167}{{\tt hep-ph/0606167}}].

\bibitem{Hou}
W.-S. Hou, {\it {Enhanced charged Higgs boson effects in $B^- \to \tau \ov
  \nu$, $\mu \ov \nu$ and $b \to \tau \ov \nu + X$}},  {\em Phys. Rev.} {\bf
  D48} (1993) 2342--2344.

\bibitem{Ake-Reck}
A.~G. Akeroyd and S.~Recksiegel, {\it {The effect of $H^\pm$ on $B^\pm \to
  \tau^\pm \nu_\tau$ and $B^\pm \to \mu^\pm \nu_\mu$}},  {\em J. Phys.} {\bf
  G29} (2003) 2311--2317, [\href{http://xxx.lanl.gov/abs/hep-ph/0306037}{{\tt
  hep-ph/0306037}}].

\bibitem{isidori-paradisi}
G.~Isidori and P.~Paradisi, {\it {Hints of large $\tan(\beta)$ in flavour
  physics}},  {\em Phys. Lett.} {\bf B639} (2006) 499--507,
  [\href{http://xxx.lanl.gov/abs/hep-ph/0605012}{{\tt hep-ph/0605012}}].

\bibitem{g-2muon}
J.~P. Miller, E.~de~Rafael, and B.~L. Roberts, {\it {Muon $g-2$: Review of
  Theory and Experiment}},  {\em Rept. Prog. Phys.} {\bf 70} (2007) 795,
  [\href{http://xxx.lanl.gov/abs/hep-ph/0703049}{{\tt hep-ph/0703049}}].

\bibitem{g-2muon2}
{\bf Muon ($g-2$)} Collaboration, G.~W. Bennett {\em et.~al.}, {\it {Final
  report of the muon E821 anomalous magnetic moment measurement at BNL}},  {\em
  Phys. Rev.} {\bf D73} (2006) 072003,
  [\href{http://xxx.lanl.gov/abs/hep-ex/0602035}{{\tt hep-ex/0602035}}].

\bibitem{g-2muon3}
K.~Hagiwara, A.~D. Martin, D.~Nomura, and T.~Teubner, {\it {Improved
  predictions for $g-2$ of the muon and $\alpha_{\rm QED}(M_Z^2)$}},
  \href{http://xxx.lanl.gov/abs/hep-ph/0611102}{{\tt hep-ph/0611102}}.

\bibitem{CHM}
M.~Czakon, U.~Haisch, and M.~Misiak, {\it {Four-loop anomalous dimensions for
  radiative flavour- changing decays}},  {\em JHEP} {\bf 03} (2007) 008,
  [\href{http://xxx.lanl.gov/abs/hep-ph/0612329}{{\tt hep-ph/0612329}}]. See
  also references therein.

\bibitem{CDGG}
M.~Ciuchini, G.~Degrassi, P.~Gambino, and G.~F. Giudice, {\it {Next-to-leading
  QCD corrections to $B \to X_s \gamma$ in Supersymmetry}},  {\em Nucl. Phys.}
  {\bf B534} (1998) 3--20, [\href{http://xxx.lanl.gov/abs/hep-ph/9806308}{{\tt
  hep-ph/9806308}}].

\bibitem{BGY}
F.~Borzumati, C.~Greub, and Y.~Yamada, {\it {Beyond leading-order corrections
  to $\bar B \to X_s \gamma$ at large $\tan\beta$: The charged-Higgs
  contribution}},  {\em Phys. Rev.} {\bf D69} (2004) 055005,
  [\href{http://xxx.lanl.gov/abs/hep-ph/0311151}{{\tt hep-ph/0311151}}].

\bibitem{DGS}
G.~Degrassi, P.~Gambino, and P.~Slavich, {\it {QCD corrections to radiative B
  decays in the MSSM with minimal flavor violation}},  {\em Phys. Lett.} {\bf
  B635} (2006) 335--342, [\href{http://xxx.lanl.gov/abs/hep-ph/0601135}{{\tt
  hep-ph/0601135}}].

\bibitem{GHM}
P.~Gambino, U.~Haisch, and M.~Misiak, {\it {Determining the sign of the $b \to
  s \gamma$ amplitude}},  {\em Phys. Rev. Lett.} {\bf 94} (2005) 061803,
  [\href{http://xxx.lanl.gov/abs/hep-ph/0410155}{{\tt hep-ph/0410155}}].

\bibitem{CERNlib}
{See the {\tt CERNlib} website: {\sf http://cernlib.web.cern.ch/cernlib/}}.

\bibitem{GonzalezMaltoni}
M.~C. Gonzalez-Garcia and M.~Maltoni, {\it {Phenomenology with Massive
  Neutrinos}},  \href{http://xxx.lanl.gov/abs/0704.1800}{{\tt 0704.1800}}.

\bibitem{BDR1}
T.~Bla{\v{z}}ek, R.~Derm{\'i}{\v{s}}ek, and S.~Raby, {\it {Predictions for
  Higgs and SUSY spectra from SO(10) Yukawa unification with $\mu > 0$}},  {\em
  Phys. Rev. Lett.} {\bf 88} (2002) 111804,
  [\href{http://xxx.lanl.gov/abs/hep-ph/0107097}{{\tt hep-ph/0107097}}].

\bibitem{BDR2}
T.~Bla{\v{z}}ek, R.~Derm{\'i}{\v{s}}ek, and S.~Raby, {\it {Yukawa unification
  in SO(10)}},  {\em Phys. Rev.} {\bf D65} (2002) 115004,
  [\href{http://xxx.lanl.gov/abs/hep-ph/0201081}{{\tt hep-ph/0201081}}].

\bibitem{Auto}
D.~Auto {\em et.~al.}, {\it {Yukawa coupling unification in supersymmetric
  models}},  {\em JHEP} {\bf 06} (2003) 023,
  [\href{http://xxx.lanl.gov/abs/hep-ph/0302155}{{\tt hep-ph/0302155}}].

\bibitem{BFPZ}
J.~A. Bagger, J.~L. Feng, N.~Polonsky, and R.-J. Zhang, {\it {Superheavy
  supersymmetry from scalar mass A-parameter fixed points}},  {\em Phys. Lett.}
  {\bf B473} (2000) 264--271,
  [\href{http://xxx.lanl.gov/abs/hep-ph/9911255}{{\tt hep-ph/9911255}}].

\bibitem{DRRR1}
R.~Derm{\'i}{\v{s}}ek, S.~Raby, L.~Roszkowski, and R.~Ruiz~de Austri, {\it
  {Dark matter and $B_s\to\mu^+ \mu^-$ with minimal SO(10) soft SUSY
  breaking}},  {\em JHEP} {\bf 04} (2003) 037,
  [\href{http://xxx.lanl.gov/abs/hep-ph/0304101}{{\tt hep-ph/0304101}}].

\bibitem{DRRR2}
R.~Derm{\'i}{\v{s}}ek, S.~Raby, L.~Roszkowski, and R.~Ruiz~de Austri, {\it
  {Dark matter and $B_s\to\mu^+\mu^-$ with minimal SO(10) soft SUSY breaking
  {II}}},  {\em JHEP} {\bf 09} (2005) 029,
  [\href{http://xxx.lanl.gov/abs/hep-ph/0507233}{{\tt hep-ph/0507233}}].

\bibitem{bsmumu-HEP07}
See talk by A. Maciel at HEP 2007, Parallel Session ``Flavour Physics and CP
  Violation'', July 20, 2007.

\bibitem{btaunu-SUSY07}
See talk by G. De Nardo at SUSY 2007, Parallel Session ``Flavour Physics'',
  July 27, 2007.

\end{thebibliography}\endgroup
